\documentclass[sn-nature,Numbered]{sn-jnl}% Style for submissions to Nature Portfolio journals
\usepackage{graphicx}%
\usepackage{multirow}%
\usepackage{amsmath,amssymb,amsfonts}%
\usepackage{amsthm}%
\usepackage{mathrsfs}%
\usepackage[title]{appendix}%
\usepackage{xcolor}%
\usepackage{textcomp}%
\usepackage{manyfoot}%
\usepackage{subcaption}
\usepackage{booktabs}%
\usepackage{diagbox}%
\usepackage{float}%
\usepackage{algorithm}%
\usepackage{ragged2e}%
\usepackage{lmodern}
\usepackage{mathrsfs}
\usepackage{algorithmicx}%
\usepackage{hyperref}
\usepackage{lineno}
\usepackage{algpseudocode}%
\usepackage{listings}%

\DeclareCaptionType{extendtable}[Extended Data Table][List of Extended Data Tables]
\DeclareCaptionType{extendfigure}[Extended Data Fig.][List of Extended 
Data Figures]

\begin{document}
%\linenumbers
\title[Article Title]{Approaching Low-Cost Cardiac Intelligence with Semi-Supervised Knowledge Distillation}
%Parameter-Efficient Semi-Supervised Learning for ECG-based Cardiovascular Disease Detection

\author[1,2]{\fnm{Rushuang} \sur{Zhou}}\email{rrushuang2-c@my.cityu.edu.hk}
\author[3,4,5]{\fnm{Yuan-Ting} \sur{Zhang}}\email{ytzhang@cuhk.edu.hk}
\author[6]{\fnm{M.Jamal} \sur{Deen}}\email{jamal@mcmaster.ca}
\author*[2,7]{\fnm{Yining} \sur{Dong}}\email{yinidong@cityu.edu.hk}
\affil[1]{\orgdiv{Department of Biomedical Engineering}, \orgname{City University of Hong Kong}, \orgaddress{\city{Hong Kong}, \country{China}}}
\affil[2]{\orgname{Hong Kong Center for Cerebro-Cardiovascular Health Engineering},\\ \orgaddress{\street{Hong Kong Science Park}, \city{Hong Kong}, \country{China}}}
\affil[3]{\orgname{Hong Kong Institutes of Medical Engineering}, \orgaddress{\city{Hong Kong}, \country{China}}}
\affil[4]{\orgdiv{Department of Electronic Engineering}, \orgname{Chinese University of Hong Kong}, \orgaddress{\city{Hong Kong}, \country{China}}}
\affil[5]{\orgname{the 
AICARE Bay Lab},\orgname{Guangdong Medical University},  \orgaddress{\city{Dong Guan}, \country{China}}}
\affil[6]{\orgdiv{Department of Electrical and Computer Engineering and School of Biomedical Engineering}, \orgname{McMaster University}, \orgaddress{\city{Hamilton, ON}, \country{Canada}}}
\affil[7]{\orgdiv{Department of Data Science}, \orgname{City University of Hong Kong}, \orgaddress{\city{Hong Kong}, \country{China}}}

\abstract{\textcolor{black}{Deploying advanced cardiac artificial intelligence for daily cardiac monitoring is hindered by its reliance on extensive medical data and high computational resources. Low-cost cardiac intelligence (LCCI) offers a promising alternative by using wearable device data, such as 1-lead electrocardiogram (ECG), but it suffers from a significant diagnostic performance gap compared to high-cost cardiac intelligence (HCCI). To bridge this gap, we propose LiteHeart, a semi-supervised knowledge distillation framework. LiteHeart introduces a region-aware distillation module to mimic how cardiologists focus on diagnostically relevant ECG regions and a cross-layer mutual information module to align the decision processes of LCCI and HCCI systems. Using a semi-supervised training strategy, LiteHeart further improves model robustness under limited supervision. Evaluated on five datasets covering over 38 cardiovascular diseases, LiteHeart substantially reduces the performance gap between LCCI and HCCI, outperforming existing methods by 4.27\% to 7.10\% in macro F1 score. These results demonstrate that LiteHeart significantly enhances the diagnostic capabilities of low-cost cardiac intelligence systems, paving the way for scalable, affordable, and accurate daily cardiac healthcare using wearable technologies. }}
%%================================%%
%% Sample for structured abstract %%
%%================================%%



%\tableofcontents
\maketitle

\section{Introduction}\label{sec:introduction}
\textcolor{black}{Cardiovascular diseases (CVDs) represent a critical global health challenge, claiming millions of lives annually and standing as the leading cause of death worldwide \cite{kelly2010promoting,mc2019cardiovascular,mendis2011global}. Compounding this crisis, a significant proportion of CVDs deaths occur out-of-hospital, alongside a vast reservoir of undiagnosed individuals, highlighting critical gaps in early detection and management \cite{zhao2019epidemiology}. Consequently, routine screening for early identification of at-risk populations and daily monitoring for patients with severe CVDs are paramount. However, implementing these vital measures at scale places immense strain on cardiologists and threatens to overwhelm public healthcare systems.} 

\begin{figure*}[t]
\begin{center}
\includegraphics[width=1\textwidth]{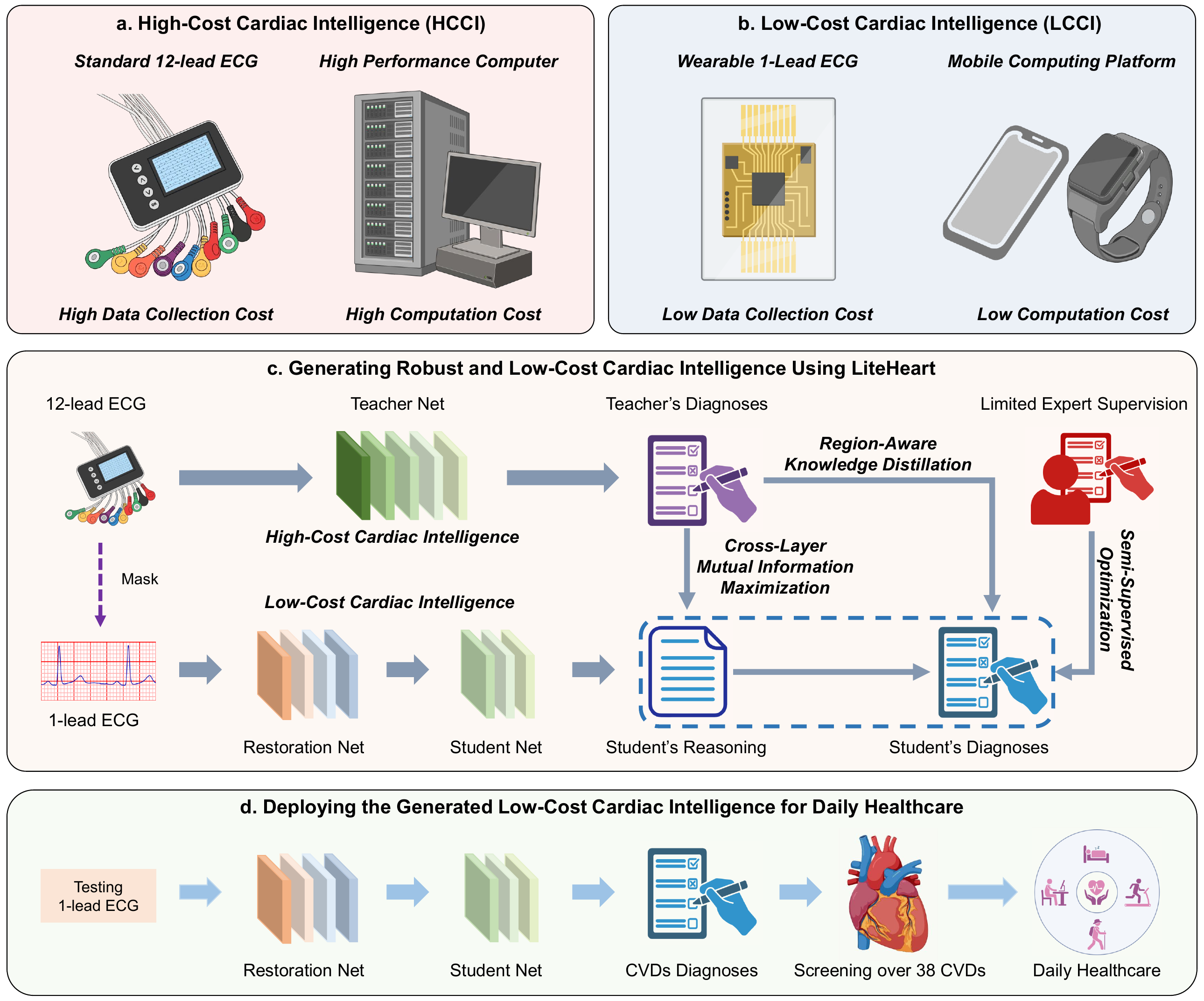}
\end{center}
\caption{\textbf{a.} High-cost cardiac intelligence utilizes 12-lead ECG signals to make CVDs diagnoses and should be deployed on high-performance computers due to its large model size.  \textbf{b.} Low-cost cardiac intelligence utilizes 1-lead ECG signals to make CVDs diagnoses and can be deployed on mobile computing platforms because of its small model size. \textbf{c.} The proposed LiteHeart introduces three modules to bridge the diagnostic performance gap between low-cost and high-cost cardiac intelligence: (1) region-aware knowledge distillation; (2) cross-layer mutual information maximization; (3) semi-supervised optimization.  \textbf{d.} Integrating the restoration net and the student net, we finally generate a robust low-cost cardiac intelligence system for daily cardiac healthcare using 1-lead ECG signals. It will first restore 12-lead ECG signals from 1-lead ECG signals and then output the corresponding CVDs diagnostic results. It can detect over 38 CVDs and show potential for daily cardiac healthcare. }
\label{fig:flowchart}%2,322,513 ECG signals from 1,558,772 patients
\end{figure*}

\textcolor{black}{Advances in deep learning offer a promising avenue through ``cardiac intelligence" – AI systems capable of automatically diagnosing CVDs from diverse inputs like 12-lead electrocardiograms (ECG) \cite{hannun2019cardiologist, ribeiro2020automatic, vaid2023foundational, al2023machine, zhou2023semi, tian2024foundation,lu2024decoding,jin2025reading, ECGFounder2025}, ultrasound \cite{ghorbani2020deep,holste2023severe,christensen2024vision} and magnetic resonance imaging \cite{fries2019weakly,jafari2023automated, wang2024screening}. Despite this potential, deploying such systems for practical, daily monitoring remains fraught with obstacles. Firstly, achieving accurate diagnoses currently demands expensive, informative medical data. While 12-lead ECG offers portability compared to MRI and ultrasound, its complex electrode placement and cumbersome data collection procedures hinder widespread, long-term use outside clinical settings. Secondly, the robust diagnostic performance of cardiac intelligence relies heavily on large-scale deep learning models. Their prohibitive computational cost and memory footprint make deployment on portable, resource-constrained devices impractical. In essence, the dual burdens of high data acquisition costs and computational demands severely limit the real-world application of current high-cost cardiac intelligence (HCCI) for daily monitoring and mass screening  (Fig.\ref{fig:flowchart}a).} 

\textcolor{black}{The rise of wearable electronics \cite{zhang2024three} presents a compelling alternative: the 1-lead ECG monitoring. This approach drastically simplifies data collection, avoiding complex electrode placement and enabling daily, long-term monitoring via consumer devices \cite{wang2024systematic}, thereby slashing data acquisition costs (Fig.\ref{fig:flowchart}b). However, the simplicity of low-cost cardiac intelligence (LCCI) using 1-lead ECG comes at a cost of significant information loss compared to the comprehensive 12-lead standard, leading to substantially degraded diagnostic performance in AI systems \cite{lai2023practical}. Knowledge distillation (KD) emerges as a potential solution, aiming to bridge this performance gap by transferring diagnostic knowledge from complex ``teacher" models (HCCI trained on 12-lead data with high computation cost) to lightweight ``student" models (LCCI targeting 1-lead data with low computation cost). \cite{hinton2015distilling,sepahvand2022novel,qin2023mvkt}.}

\textcolor{black}{While promising, LCCI built on 1-lead ECG and knowledge distillation is still in its infancy. Even state-of-the-art (SOTA) methods exhibit a substantial performance deficit compared to HCCI, which can be attributed to four key factors. (1) Despite the assistance of 12-lead ECG signals during the knowledge transfer process, student models ultimately rely solely on limited disease patterns from 1-lead ECG for diagnosis, imposing a fundamental performance ceiling \cite{lai2023practical}. (2) Cardiologists diagnose CVDs by identifying and analyzing specific Regions of Interests (ROIs) within the ECG signals that reveal disease patterns. However, existing knowledge distillation methods fail to adequately transfer this crucial ROI-based diagnostic process to the student model, resulting in poor CVDs diagnostic performance. (3) Small-scale student models inherently possess weaker learning capabilities than large teacher models \cite{huang2022knowledge}. Standard knowledge distillation methods, which align outputs or specific layer activations, often struggle when there is a vast capacity gap between student and teacher models, leading to suboptimal knowledge transfer efficiency. (4) The high cost of expert ECG annotation creates a scarcity of labeled data in the downstream dataset, which limits the utility of existing knowledge distillation methods that require substantial labeled data to effectively guide the student model, presenting a major barrier for cardiac intelligence deployment\cite{phuong2019towards,hao2023revisit}.}

\textcolor{black}{To overcome these limitations and realize truly practical low-cost cardiac intelligence, we developed a novel semi-supervised knowledge distillation framework (LiteHeart, Fig.\ref{fig:flowchart}c) that consists of three critical modules: (1) a region-aware knowledge distillation module to explicitly transfer the cardiologist's ROI-based diagnostic process, (2) a cross-layer mutual information maximization module to address the capacity mismatch between large-scale teachers and small-scale students, and (3) a semi-supervised optimization module to leverage the abundant unlabeled ECG data in clinical settings. For daily health care deployment, the 12-lead ECG restoration net is integrated with the student net to achieve robust diagnostic performance while offering high inference efficiency, minimal data collection burden, and reduced reliance on costly expert annotations (Fig.\ref{fig:flowchart}d).}

\textcolor{black}{The experimental results on five ECG datasets demonstrate the superior performance of the proposed LiteHeart framework against the SOTA solutions. Specifically, LiteHeart reduces the diagnostic performance gap between LCCI and HCCI almost in half and surpasses its best competitor by 4.27\% to 7.10\% on the macro F1 score. Compared with HCCI using 12-lead ECG signals, our proposed LCCI can reduce inference latency and memory consumption by 2.09 and 14.82 times. This combination of high accuracy and minimal resource requirements makes the LiteHeart-generated system well-suited for deployment on wearable devices and mobile platforms, enabling fast, accurate, and truly accessible CVDs screening outside the hospital. In summary, LiteHeart provides a robust platform for developing deployable low-cost cardiac intelligence, paving the way for accessible routine CVDs screening and daily cardiac health monitoring via wearable devices. As such, it demonstrates potential for the early identification of at-risk populations through daily monitoring, and might contribute to reducing the proportion of CVDs deaths occurring out-of-hospital through timely alarm.}
\section{Results}\label{sec:result}
\subsection{Overview}
In the following, we provide an overview of the proposed LiteHeart, which is described in Section \ref{sec:method} in detail. (1, Extended Data Fig.\ref{fig:flowchart_details}a) Region-aware knowledge distillation to transfer the cardiologist's ROI-based diagnostic process. First, with great learning capacity, the teacher net acts as the HCCI that is optimized to make accurate diagnoses using 12-lead ECG signals. Concurrently, a restoration net recovers an approximation of the 12-lead signal from the 1-lead ECG input, enriching the information available for a lightweight student net to perform CVDs diagnosis. The core innovation involves randomly distorting regions within both the real and restored 12-lead signals. The key insight is that distortion of clinically critical regions would alter the diagnosis, while distortion of other regions would not. We simulate the cardiologist feedback using the teacher net's predictions, and then train the student net to mimic its reactions. By aligning with the teacher's sensitivity to specific signal regions, the student learns where to look (identifying ROIs) and how disease manifests spatially, directly emulating clinical reasoning without requiring direct cardiologist annotations on distortions. (2, Extended Data Fig.\ref{fig:flowchart_details}b) Cross-layer mutual information maximization to strengthen the relationships between the student's reasoning and the teacher's diagnosis. Moving beyond simple output alignment, we introduce a discrimination net to evaluate whether the student's reasoning process is consistent with the teacher's final diagnosis. By minimizing the discrimination loss, we effectively maximize the mutual information between the student's evolving understanding and the teacher's diagnostic conclusion. This deeper alignment, focusing on the reasoning pathway rather than just the endpoint, significantly boosts distillation efficiency and final performance. (3, Extended Data Fig.\ref{fig:flowchart_details}c) Semi-supervised optimization. To overcome the label scarcity problem in clinical practice, we utilize the information from massive unlabeled data through a simple yet effective semi-supervised optimization pipeline. The objective function is formulated to effectively utilize both labeled and unlabeled samples \cite{phuong2019towards,hao2023revisit}, enabling effective student training even when expert annotations are limited.

\subsection{Datasets and experiment protocols}
First of all, we pre-train the restoration net and the teacher net on a large-scale ECG dataset with 2,322,513 12-lead ECG signals \cite{ribeiro2019tele,ribeiro2020automatic}. Then, the teacher net and the restoration net are fine-tuned on the downstream datasets. The implementation details of the pertaining and the fine-tuning process are presented in Section \ref{sec:pre} and the Extended Data Fig.\ref{fig:flowchart_details}d-e. We use five downstream datasets to evaluate the performance of the low-cost cardiac intelligence generated by the proposed LiteHeart framework and other competitors. Above all, a wearable 12-lead ECG signals (SMU) dataset provided by Southern Medical University is used for evaluating the model performance on mobile CVDs screening tasks \cite{lai2023practical}. Specifically, 7000 wearable ECG signals from the dataset are publicly available and thus used for our experiments. Due to the scarcity of wearable ECG datasets that cover a broad range of CVDs, we also include clinical ECG datasets for benchmarking. Specifically, 12-lead ECG signals from the Georgia 12-lead ECG signals Challenge (G12EC) dataset \cite{alday2020classification}, the Physikalisch-Technische Bundesanstalt (PTB-XL) dataset \cite{wagner2020ptb}, the Ningbo dataset \cite{zheng2020optimal} and the Chapman-Shaoxing dataset \cite{zheng202012} are collected in a hospital or clinical setting \cite{alday2020classification,reyna2021will}. Specifically, the G12EC database contains 10344 ECG signals from 10,344 people in Atlanta, Georgia, USA, and the PTB-XL database comprises 21837 signals from 18885 patients in Brunswick, Germany. The Chapman database contains 10,646 signals from 10646 patients in Shaoxing, Zhejiang, China, and the Ningbo database encompasses 40258 signals from 40258 patients in Ningbo, Zhejiang, China. Only 34,905 signals in the Ningbo database are publicly available \cite {alday2020classification}. The signals from the above downstream databases are around 10-15 seconds with various sampling rates. For signal pre-processing, a band-pass filter (1-47Hz) is applied to remove the power-line interference and baseline drift. Then, the pre-processed signals are normalized using z-score normalization. Here, 1-lead ECG signals are generated by masking out the remaining leads of 12-lead ECG signals except lead I. Additionally, there are 38 different CVDs within the five downstream datasets and each database contains over 17 different CVDs. The CVDs analyzed in our study and their abbreviations can be found in the Extended Data Table \ref{tab:anotation}. Note that multiple CVDs can be identified from one ECG segment simultaneously. Consequently, CVDs diagnosis using ECG can be formulated as a multi-label classification task.

For model training and evaluation, we divide a given downstream dataset into a training set $\mathcal{D}$, a validation set $\mathcal{D}_V$, and a test set $\mathcal{D}_{T}$, in a ratio of 8: 1: 1. Subsequently, the training set is divided into a labeled training set $\mathcal{D}_B$ and an unlabeled training set $\mathcal{D}_{U}$ in a ratio of 1: 9. In other words, 10\% training samples are labeled while the remaining 90\% are unlabeled.  The restoration net is fine-tuned on the labeled training set and the unlabeled training set. The teacher net is only fine-tuned on the training set with labels. Based on the proposed LiteHeart framework, the low-cost cardiac intelligence is generated using the labeled and unlabeled sets and is evaluated on the test set using four metrics: macro F1 score, macro $F_{\beta=2}$, mean average precision (MAP), and macro AUC. They measure the average performance of the evaluated model across different CVDs and range from 0\% to 100\%. For example, the macro F1 score is the average F1 score across various types of CVDs. The value of the parameter $\beta$ is 2 for all the corresponding experiments as recommended by \cite{strodthoff2020deep}. To ensure reproducibility, all the experiments are repeated four times using different random seeds.  The mean value and the standard deviation of the four metrics are used for analysis. A detailed computation process for the metrics can be found in \cite{zhou2024computation}. The validation set is used to terminate the training process if the validation error does not decrease for 10 epochs.
\begin{figure*}[t]
\begin{center}
\includegraphics[width=1\textwidth]{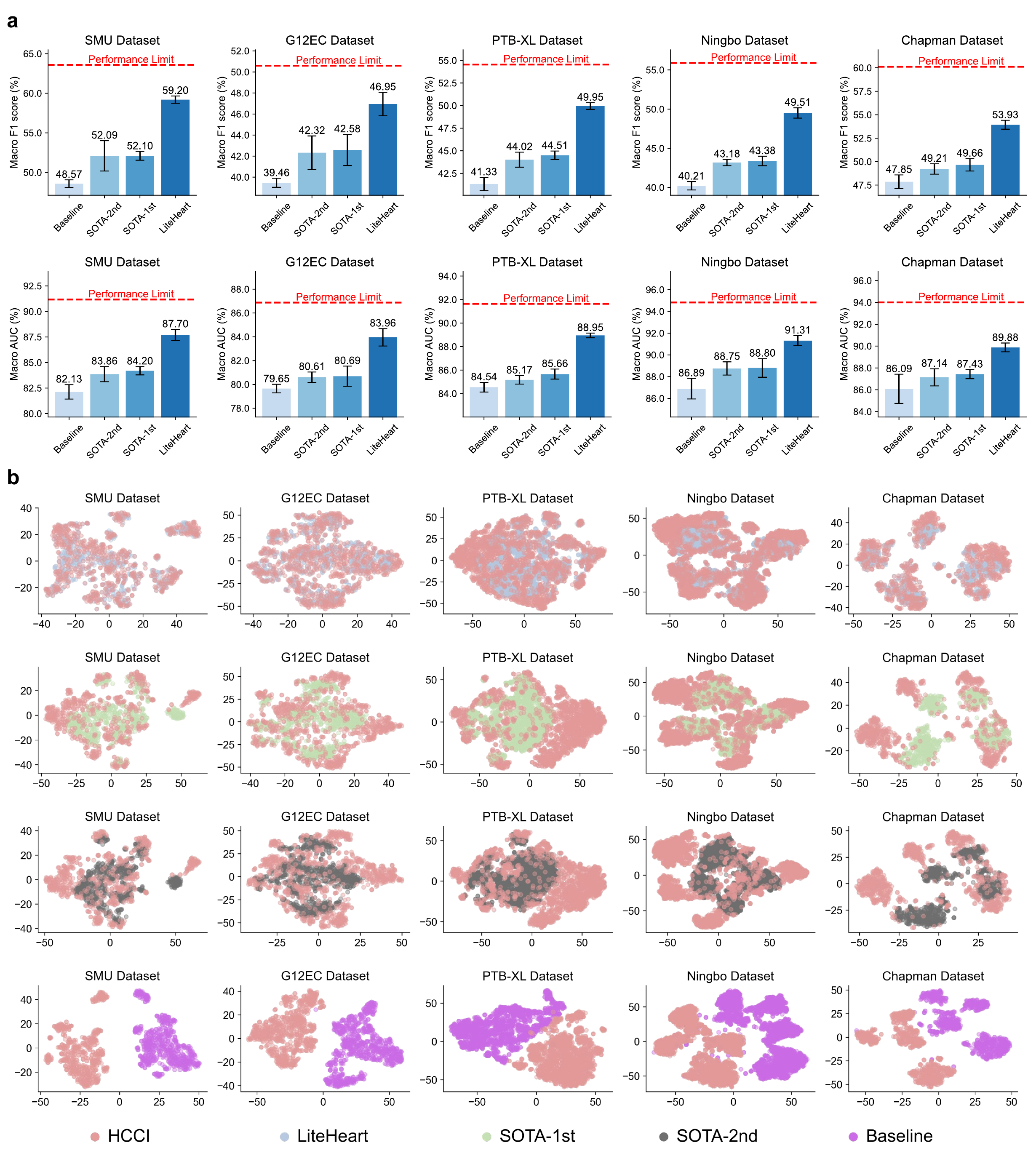}
\end{center}
\caption{\textbf{a.} CVDs diagnostic performance of the low-cost cardiac intelligence generated by the proposed LiteHeart and the SOTA methods on different downstream datasets. The red dashed lines denote the performance of the high-cost teacher nets using 12-lead ECG signals, which serve as the performance ceiling of low-cost cardiac intelligence. The baseline of low-cost cardiac intelligence is formulated by deactivating the knowledge distillation. \textbf{b.} T-SNE visualization of the last layer logits output by the high-cost cardiac intelligence, the low-cost cardiac intelligence generated by LiteHeart, SOTA-1st, SOTA-2nd, and the baseline method. The circles with different colors describe the distribution of CVDs diagnoses made by different models.  }
\label{fig:main_results}
\end{figure*}

\subsection{Diagnostic performance}
Here, we reproduce diverse SOTA methods in knowledge distillation for benchmarking, including vanilla knowledge distillation (KD) \cite{hinton2015distilling}, decoupled knowledge distillation (DKD) \cite{zhao2022decoupled}, correlation congruence for knowledge distillation (CC) \cite{peng2019correlation}, relational knowledge distillation (RKD) \cite{park2019relational}, DIST \cite{huang2022knowledge}, probabilistic knowledge transfer (PKT) \cite{passalis2020probabilistic}, partial softmax knowledge distillation (PSM) \cite{song2021handling}, Wasserstein knowledge distillation (WKD) \cite{lv2024wasserstein}, and multi-view knowledge transferring (MVKT) \cite{qin2023mvkt}. We ensure that all the compared methods share the same architectures for the teacher and the student nets. The teacher net has 50.5 million parameters, while the student net has 1.6 million parameters, respectively. Details about their architectures are provided in the Supplementary Materials. As shown in Fig.\ref{fig:main_results}a, we present the diagnostic performance of the low-cost cardiac intelligence generated by LiteHeart, the best and second best SOTA methods. Detailed performance of all the compared methods is provided in the Extended Data Table \ref{tab:compare}. According to the results, we observe that LiteHeart reduces the performance gap between low-cost and high-cost cardiac intelligence on different datasets almost by half. For example, on the SMU dataset, LiteHeart decreases its gap in the macro F1 score from 15.01\% to 4.39\%. Compared with the SOTA methods in knowledge distillation, LiteHeart demonstrates superior performance on all datasets and evaluation metrics. For example, LiteHeart outperforms the SOTA-1st by 4.27\% to 7.10\% on the macro F1 score calculated on the five datasets. Utilizing the T-SNE technique \cite{van2008visualizing}, we visualize the features extracted from the last layers of high-cost cardiac intelligence and the low-cost cardiac intelligence generated by various methods. The discrepancy between the feature distributions obtained from high-cost and low-cost cardiac intelligence reflects their differences in diagnosing CVDs. As illustrated in Fig.\ref{fig:main_results}b, LiteHeart significantly outperforms SOTA-1st in minimizing this discrepancy, providing a straightforward demonstration of its superiority in helping low-cost cardiac intelligence mimic the diagnostic process of high-cost cardiac intelligence.
\begin{figure*}[t]
\begin{center}
\includegraphics[width=1\textwidth]{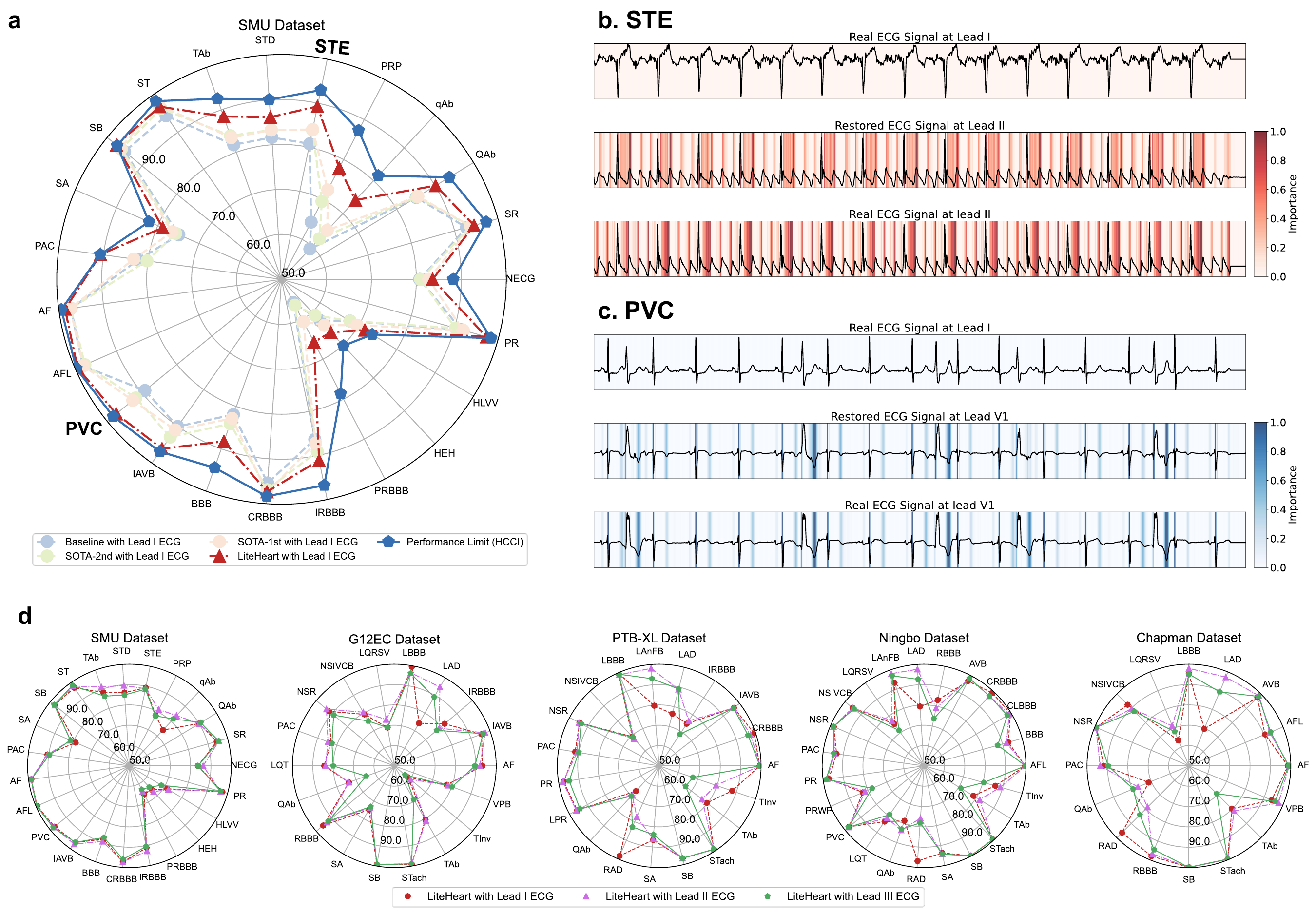}
\end{center}
\caption{\textbf{a}. AUC of different models on various CVDs from the SMU dataset. \textbf{b}-\textbf{c}. Visualization of ROIs for diagnosing “lambda-wave” ST-elevation (\textbf{STE}) and premature ventricular contractions (\textbf{PVC}) using the Grad-GAM approach. For example, ST-elevation is difficult to diagnose using the real lead I ECG signal. With the restoration net, a restored 12-lead ECG signal can be generated using the lead I signal, and the corresponding diagnostic patterns are recovered. As shown in lead II of the restored signal, the ROIs for diagnosing “lambda-wave” ST-elevation are highlighted using the low-cost cardiac intelligence generated by LiteHeart. For comparison, we visualize the lead II of the real ECG signal and highlight the ROIs using high-cost cardiac intelligence. \textbf{d.} The impact of ECG lead selection on the diagnostic performance of the low-cost cardiac intelligence.}
\label{fig:gradcam}
\end{figure*}

In Fig.\ref{fig:gradcam}a, we present the AUC of the low-cost cardiac intelligence generated by LiteHeart on each CVD using the SMU dataset. In the Extended Data Fig. \ref{fig:radar_extend}, we present the corresponding results on the G12EC, PTB-XL, Ningbo, and Chapman datasets. The results indicated that LiteHeart enhances the performance of low-cost cardiac intelligence for identifying CVDs that are challenging to diagnose using lead I ECG signals. For example, LiteHeart bridges the AUC gap between low-cost and high-cost cardiac intelligence in diagnosing premature ventricular contractions (PVC),  ST-elevation (STE), etc. The features of PVC and STE mainly exist in lead V1 and lead II, respectively, making them challenging to diagnose using only lead I ECG signals. As shown in Fig.\ref{fig:gradcam}b and Fig.\ref{fig:gradcam}c, we observe that LiteHeart's restoration net successfully recovers the critical features for diagnosing PVC in lead V1 and STE in lead II, thereby overcoming the information bottleneck of a single lead I ECG. Subsequently, we use the Grad-CAM approach \cite{selvaraju2017grad} to visualize the ROIs identified by low-cost and high-cost cardiac intelligence during their diagnostic process, as shown in Fig.\ref{fig:gradcam}b and Fig.\ref{fig:gradcam}c. With the assistance of LiteHeart, low-cost cardiac intelligence successfully identifies the correct ROIs for diagnosing PVC and STE, suggesting that it shares a similar diagnostic process with high-cost cardiac intelligence. The above observations demonstrate the superiority and effectiveness of the proposed LiteHeart in increasing the CVDs diagnostic performance of low-cost cardiac intelligence. 

We further evaluate the performance of LiteHeart using different 1-lead ECG signals and present the results in Fig.\ref{fig:gradcam}d. It can be observed that the choice of ECG leads influences the diagnostic performance of low-cost cardiac intelligence. For example, using lead I ECG signals, the model demonstrates robust performance in diagnosing right axis deviation (RAD) but struggles at diagnosing left axis deviation (LAD). On the contrary, using lead II and lead III ECG signals will enable the model to diagnose LAD but weaken its ability to diagnose RAD. For different CVDs, the leads that provide critical information for diagnosis will contribute most to the diagnostic results of cardiologists, which explains the above phenomenon. This suggests that we should customize the lead settings for patients with different needs for cardiac healthcare and CVDs screening.  

At the same time, we reduce the ratio of labeled samples with each downstream dataset from 10\% to 5\% and present the diagnostic performance of the low-cost cardiac intelligence generated by different methods in Extended Data Fig.\ref{fig:compare_tiny_005}. Detailed performance of all the compared methods is provided in the Extended Data Table \ref{tab:compare_005}. This experiment poses a great challenge to the knowledge distillation performance under extremely scarce supervised information. The results demonstrate that the performance of the proposed LiteHeart compared to the SOTA methods remains superior in this challenging setting. For example, it outperforms the best-performing SOTA by 6.7\% to 10.71\% on the macro F1 score calculated on the five datasets, which demonstrates its robustness in label scarcity scenarios.

\subsection{Inference efficiency}
\begin{figure*}[t]
\begin{center}
\includegraphics[width=1\textwidth]{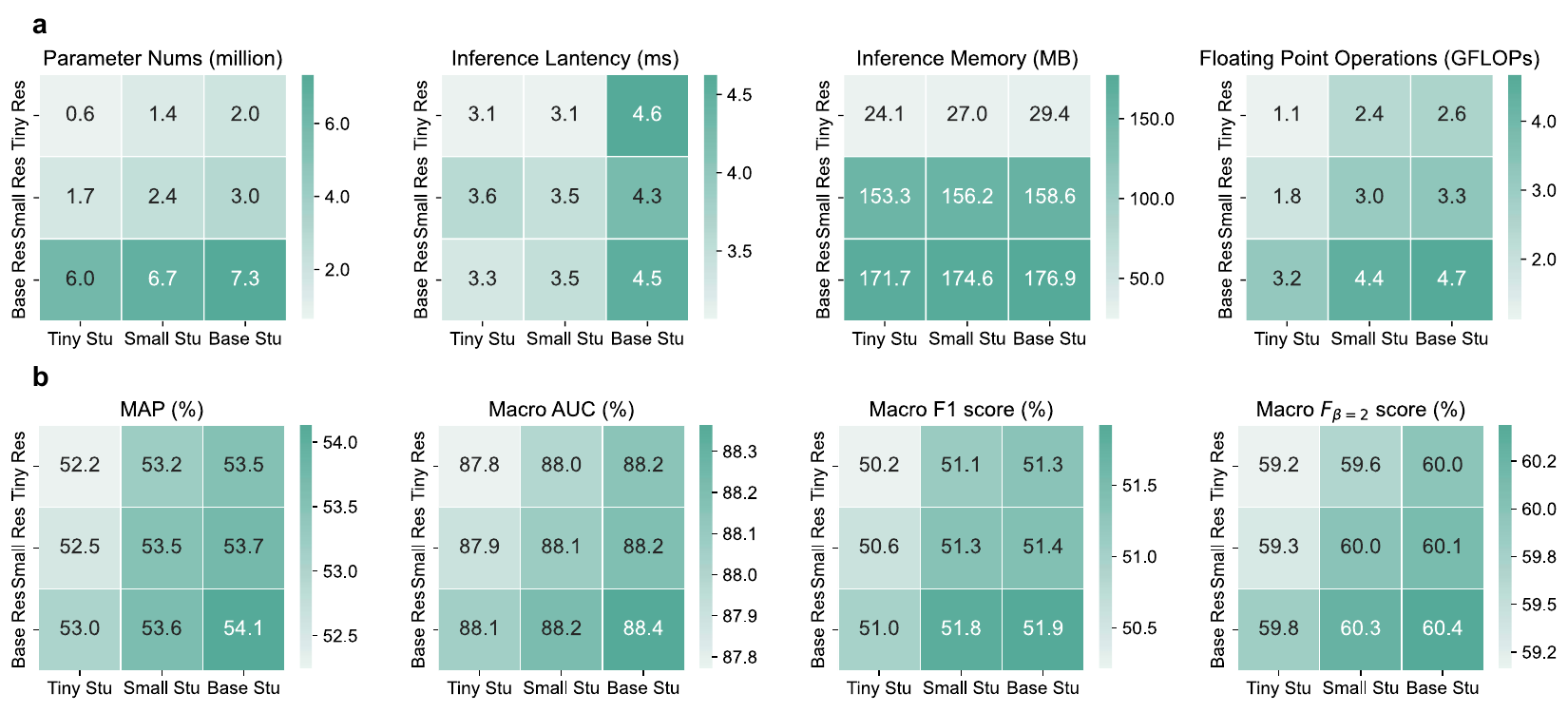}
\end{center}
\caption{\textbf{a}. The impact of restoration net and student net sizes on the inference efficiency of the low-cost cardiac intelligence generated by LiteHeart. \textbf{b}. The impact of restoration net and student net sizes on the CVDs diagnostic performance. Here, ‘Res’ and 'Stu' are the abbreviations of the restoration net and the student net, respectively. Specifically, a tiny student net has 0.26 million parameters, and a tiny restoration net has 0.36 million parameters. A small student net has 1.01 million parameters, and a small restoration net has 1.43 million parameters. A base student net has 1.60 million parameters, and a base restoration net has 5.71 million parameters.} 
\label{fig:efficiency}
\end{figure*}

Apart from low data collection costs, another objective of low-cost cardiac intelligence is to achieve high computational efficiency during model inference. In our study, the low-cost cardiac intelligence consists of a restoration net for 12-lead ECG signals restoration and a student net for CVDs diagnosis, which have 5.71 million and 1.60 million parameters, respectively. From the standpoint of practical applications, the budget of computational resources varies across different hardware platforms. For portable devices with limited processing units,  the sizes of the restoration net and student net should be compressed to achieve a satisfying inference efficiency. At the same time, the generated low-cost cardiac intelligence should maintain a robust diagnostic performance under a tight budget in model learning capacity. In response, we adjust the sizes of the restoration net and the student net and evaluate the inference efficiency and diagnostic performance of the generated low-cost cardiac intelligence by LiteHeart. Specifically, we quantify the model learning capacity using the number of parameters. Under a batch size of 4, we compute the inference latency, the inference memory usage, and the floating point operations to evaluate the computational efficiency during model inference. Additionally, we calculate the averaged macro F1 score, macro $F_{\beta=2}$ score, and the macro AUC across five downstream datasets. Note that 10\% of the training samples were labeled. As shown in Fig.\ref{fig:efficiency}a-b, the low-cost cardiac intelligence generated by the proposed LiteHeart can maintain its diagnostic performance under limited computational resources. Specifically, the fluctuations of macro F1 score and macro AUC are only 1.7\% and 0.6\% when the model learning capacity gradually reduces from 7.3 million parameters to 0.6 million parameters. Additionally, the inference latency, the inference memory usage, and the floating point operations of the generated low-cost cardiac intelligence decrease from 4.5 ms to 3.1 ms, 176.9 MB to 24.1 MB and 4.7 GFLOPs to 1.1 GFLOPs. In contrast, the high-cost cardiac intelligence formulated by the teacher net has an inference latency of 6.48 ms, memory usage of 357.2 MB, and requires 39.45 GFLOPs. It can be observed that low-cost cardiac intelligence can achieve a remarkable 2.09 times increase in inference speed, a 14.82 times reduction in memory usage, and a 35.86 times decrease in floating point operations when compared to high-cost cardiac intelligence. \textcolor{black}{Furthermore, the results indicate that the capacity of the student net plays a more important role than that of the restoration net in enhancing diagnostic performance. For instance, the diagnostic system with a tiny restoration net and a base student net achieves a higher mean average precision (MAP) compared to the system with a base restoration net and a tiny student net.} In summary, the superior inference efficiency of low-cost cardiac intelligence makes it suitable for deployment on portable devices with limited computational resources. Moreover, its robust diagnostic performance across varying model capacities underscores its stability in clinical practice.  

\begin{figure*}[t]
\begin{center}
\includegraphics[width=1\textwidth]{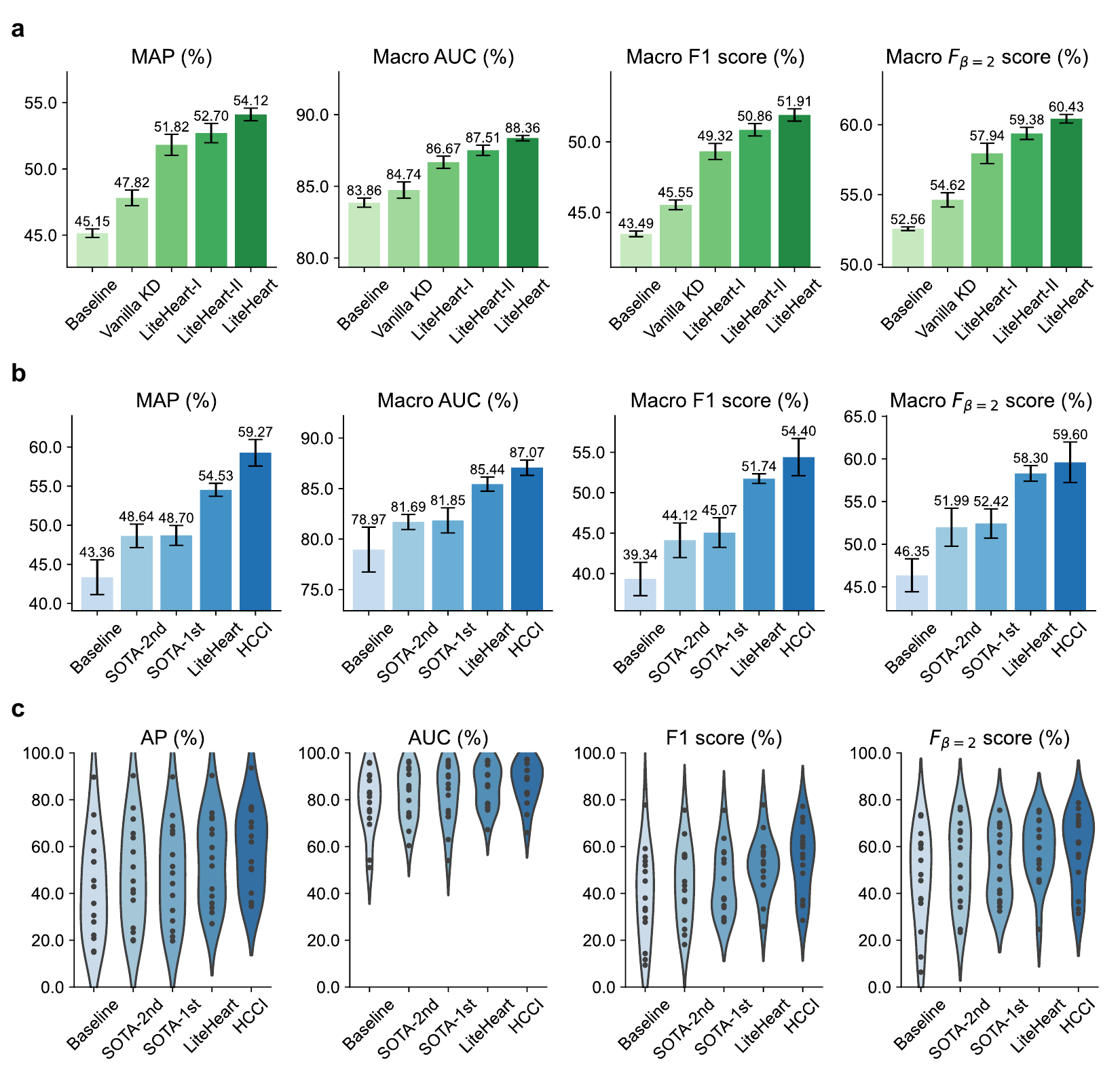}
\end{center}
\caption{\textbf{a.} Ablation study of the proposed LiteHeart. The baseline of low-cost cardiac intelligence is formulated by deactivating the knowledge distillation. \textbf{b.} Cross-device external validation of high-cost cardiac intelligence (HCCI) and the low-cost cardiac intelligence generated by different methods. The mean value and the standard deviations of all metrics are calculated across four random seeds. \textbf{c.} The distribution of the diagnostic performance across all evaluated CVDs. For example, a point within the violin plot denotes the AUC of one CVD. The centroid of the points denotes the value of macro AUC for CVDs detection.   }
\label{fig:other_results}
\end{figure*}

\subsection{Ablation study}
To provide insight into the contribution of different modules in LiteHeart, we successively add them to the vanilla knowledge distillation \cite{hinton2015distilling} and present the diagnostic performance of the generated low-cost cardiac intelligence in Fig.\ref{fig:other_results}a. Specifically, LiteHeart-I is constructed by replacing the vanilla knowledge distillation with the proposed region-aware knowledge distillation. LiteHeart-II is created by adding the cross-layer mutual information maximization module to LiteHeart-I. Compared with LiteHeart-II, LiteHeart includes unlabeled samples for semi-supervised optimization. For simplicity, we calculate the average performance across five downstream datasets. Its standard deviations across four random seeds are shown as error bars. Note that 10\% of the training samples were labeled. Additionally, we use a base student net with 1.60 million parameters and a base restoration net with 5.71 million parameters for experiments. First, the results demonstrate that the region-aware knowledge distillation method significantly improves the diagnostic performance of low-cost cardiac intelligence. For instance, the macro F1 score and MAP increase by 3.77\% and 4\% when the module is activated (LiteHeart-I vs Vanilla KD). Second, it can be observed that introducing the cross-layer mutual information maximization module enhances the knowledge distillation performance. For example, the macro F1 score and macro AUC increase by 1.54\% and 0.84\% when the module is added (LiteHeart-II vs LiteHeart-I). Third, utilizing the information within the unlabeled data in a semi-supervised manner benefits the diagnostic performance under limited supervision. Specifically, on the Chapman dataset, the MAP and the macro $F_{\beta=2}$ score increase by 1.42\% and 1.05\% when the unlabeled samples are included. 

\subsection{Cross-device external validation}
External validation has been widely used to evaluate the generalization performance of the CVDs diagnostic systems on unseen datasets \cite{lai2023practical,vaid2023foundational}.  It assumes that the training datasets and the unseen external datasets are collected using similar devices, ensuring that the ECG signals sampled from them follow an identical distribution. As introduced in our study, one application of low-cost cardiac intelligence is mobile cardiac healthcare using wearable devices. Unlike clinical ECG devices, the development of wearable devices is still in its infancy, and they usually have immature hardware designs and diverse electrode settings. Thus, collecting large scale ECG signals using wearable devices and dividing them into training sets and unseen external datasets is difficult. In our study, we propose a practical cross-device external validation approach, which sets large-scale clinical ECG datasets (G12EC, PTB-XL, Ningbo, and Chapman) and small-scale wearable ECG datasets (SMU) as training sets and the unseen external dataset, respectively. The CVDs that co-exist in the unseen external dataset and the training sets are used for evaluation, including NSR, QAb, TAb, IAVB, BBB, CRBBB, IRBBB, CLBBB, SB, SA, PAC, AF, AFL, PVC, and PR. We set the ratio of labeled samples within the training sets to 10\% and present the diagnostic performance of the low-cost cardiac intelligence generated by LiteHeart, the best and the second best SOTA methods in Fig.\ref{fig:other_results}b-c. Here, wearable 1-lead ECG signals for external validation are generated by masking out the remaining leads of wearable 12-lead ECG signals except lead I. The diagnostic performance of high-cost cardiac intelligence is also reported for comparison. Detailed performance of all the compared methods is provided in the Extended Data Table \ref{tab:external_tiny10}. The results highlight the superior generalization performance of the low-cost cardiac intelligence generated by the proposed LiteHeart in the cross-device external validation. When the ratio of labeled samples is 10\%, LiteHeart outperforms the best competitor SOTA-1st by 3.59\% and 5.88\% on the macro AUC and the macro $F_{\beta=2}$ score. 

Additionally, we reduce the ratio of labeled samples within the training set from 10\% to 2.5\% and present the performance of different models in Extended Data Fig.\ref{fig:external_extend}. 
LiteHeart consistently outperforms all competitors. At a 5\% labeling ratio, LiteHeart surpasses the best competitor by 6.02\% in MAP and 5.65\% in macro F1 score. With only 2.5\% labeled data, LiteHeart outperforms the best competitor by 5.43\% in macro F1 score and 5.33\% in macro $F_{\beta=2}$ score. Critically, LiteHeart demonstrates a remarkable ability to bridge the performance gap between low-cost and high-cost cardiac intelligence in cross-device external validation. At 10\% labels, the macro AUC gap is reduced from 8.10\% to 1.63\% and the macro F1 gap is reduced from 15.06\% to 2.66\%. When the ratio of labeled samples decreases to 5\%, the MAP gap is reduced from 18.63\% to 6.56\% and the $F_{\beta=2}$ gap is reduced from 12.49\% to 0.24\%. With only 2.5\% labeled data, the gaps in macro AUC and macro F1 score are reduced from 10.21\% to 1.81\% and from 15.20\% to 4.17\%. Detailed performance of all compared methods is provided in Extended Data Table \ref{tab:external_tiny05} and Extended Data Table \ref{tab:external_tiny025}.

\section{Discussion}\label{sec:discussion}
\textcolor{black}{LiteHeart is proposed to generate low-cost cardiovascular diseases (CVDs) diagnostic systems with high inference efficiency, reduced data collection and computational costs. By integrating a region-aware knowledge distillation module, a cross-layer mutual information maximization module, and a semi-supervised optimization module, the low-cost cardiac intelligence generated by LiteHeart achieves superior performance across diverse, multi-center ECG datasets. Compared to traditional methods, LiteHeart significantly narrows the diagnostic performance gap between low-cost and high-cost cardiac intelligence (Fig.\ref{fig:main_results}a, Fig.\ref{fig:other_results}c, Extended Data Fig.\ref{fig:compare_tiny_005}a-b, Extended Data Table \ref{tab:compare}, Extended Data Table \ref{tab:compare_005}), making it promising for wearable device deployment.}

\textcolor{black}{Cardiovascular diseases, the global leading cause of death \cite{kelly2010promoting,mc2019cardiovascular,mendis2011global}, demand scalable solutions for early detection and daily monitoring. While wearable 1-lead ECG offers a practical pathway for continuous monitoring \cite{zhang2024three, wang2024systematic}, directly making CVDs diagnosis using 1-lead ECG on portable devices leads to inferior diagnostic performance due to inherent information loss and constrained learning capacity. LiteHeart successfully overcomes this issue by introducing a unified framework to improve CVDs diagnostic performance on portable devices. As evidenced by Fig.\ref{fig:gradcam}, LiteHeart achieves robust performance even for CVDs that are notoriously difficult to diagnose from 1-lead signals, such as Premature Ventricular Contractions (PVC) and ST-Elevation (STE). Additionally, LiteHeart delivers exceptional efficiency essential for wearable deployment. The integrated system (Fig.\ref{fig:flowchart}d) achieves a 14.82 times reduction in memory consumption and is 2.09 times faster than high-cost cardiac intelligence. Crucially, its diagnostic performance remains stable across computational budgets (Fig.\ref{fig:efficiency}). This demonstrates that the proposed LiteHeart can adapt well to devices with various computational resources, such as mobile phones and smartwatches, without noticeable trade-offs in accuracy.}

\textcolor{black}{Except for ECG and cardiac healthcare, LiteHeart might also benefit the research in wearable physiological signals, which are used for sleep stage recognition\cite{memar2017novel,eldele2021attention}, mental disorders diagnosis\cite{yasin2021eeg, baygin2021automated} and epilepsy detection\cite{zhang2025daff, sun2025multi}. Multi-channel physiological signals, such as electroencephalography (EEG) and electromyography (EMG), are the gold standard for diagnosing various mental and physical diseases in clinical practice. However, similar to 12-lead ECG, such signals are difficult to collect outside the hospital and medical laboratory due to the complexity of electrode placements, making it challenging for daily healthcare applications. Fortunately, the information restoration and region-aware distillation pipeline used in the proposed LiteHeart can provide insights into how to achieve comparable diagnostic performance using wearable single-channel signals under limited expert supervision. Its effectiveness on low-cost CVDs diagnoses indicates its potential for extending precise mental and physical healthcare from hospitals to daily life, which deserves further empirical validation. }

\textcolor{black}{Furthermore, LiteHeart might contribute to the broader machine intelligence domains by introducing a promising solution to effectively transfer the knowledge from large-scale foundation models to small-scale student models, which shows potential for promoting the development of embodied and on-edge intelligence. Despite the powerful performance of modern artificial intelligence, its applications on edge and mobile devices are limited by the high computational costs induced by large model sizes. Knowledge distillation\cite{hinton2015distilling} provides a feasible solution for model compression, but its effectiveness is often limited by the learning capacity gap between teachers and students. LiteHeart addresses this by maximizing the mutual information between the student's reasoning process and the teacher's outputs, ensuring a more comprehensive and effective knowledge transfer. Ablation studies on five downstream datasets demonstrate that the proposed technique is able to boost the knowledge distillation efficiency and improve the student's performance (Fig.\ref{fig:other_results}a). 
Additionally, LiteHeart offers a new perspective on learning with incomplete data. Moving beyond traditional methods that simply restore missing values, LiteHeart uses region-aware knowledge distillation to ensure the model maintains a consistent, ROI-based inference process when using restored and complete true data. Its success with 1-lead ECG signals holds promise for inspiring future research aimed at addressing other real-world problems.}

While LiteHeart represents a major advance, we have to acknowledge that it does not entirely eliminate the performance gap. Enhancing the 12-lead restoration using advanced generative models is a promising direction, such as diffusion models \cite{bedin2024leveraging}, generative adversarial models \cite{joo2023twelve}, and visual auto-regressive models \cite{tian2024visual}. However, the computational burden of current state-of-the-art generative models poses a challenge for low-cost deployment. Future work could therefore focus on developing efficient generative architectures or distillation techniques tailored for restoration within the LiteHeart paradigm, carefully balancing fidelity gains with computational overhead.

To summarize the discussion, we can conclude that LiteHeart provides a robust platform for developing deployable low-cost cardiac intelligence, paving the way for accessible routine CVDs screening and daily cardiac health monitoring via wearable devices. Beyond cardiovascular diseases detection, it demonstrates potentials for generalizing to the broader digital healthcare and machine intelligence domains. 

\section{Methods}\label{sec:method}
\subsection{Preliminaries about knowledge distillation}
Knowledge distillation has shown the potential to be able to transfer the knowledge from 12-lead ECG signals to 1-lead ECG signals \cite{qin2023mvkt, hinton2015distilling}. A common practice of this technique for multi-label classification can be formulated as, 
\begin{equation}
\label{Eq:knowledge}
\mathcal{L}_{K}(\mathcal{D}_B,\theta_s,\theta_t)=-\frac{\tau^2}{NC}\sum^{N}_{i=1}\sum^{C}_{c=1}(1-\sigma(p_{i,c}^{t}/\tau))\log(1-\sigma(p_{i,c}^s/\tau))+\sigma(p_{i,c}^{t}/\tau)\log \sigma(p_{i,c}^{s}/\tau),
\end{equation}
\begin{equation}
\label{Eq:supervised_loss_pre}
\mathcal{L}_{Y}(\mathcal{D}_B,\theta_s)=-\frac{1}{NC}\sum^{N}_{i=1}\sum^{C}_{c=1}(1-y_{i,c})\log(1-\sigma(p_{i,c}^s))+y_{i,c}\log \sigma(p_{i,c}^{s}),
\end{equation}
\begin{equation}
\label{Eq:total_loss_vanilla}
\mathcal{L} = \mathcal{L}_{Y}(\mathcal{D}_B,\theta_s) + \alpha\mathcal{L}_{K}(\mathcal{D}_B,\theta_s,\theta_t).
\end{equation}
where $\mathcal{L}_{K}$ is the knowledge distillation loss and $\mathcal{L}_{Y}$ is the multi-label classification loss for the student net. $p_{i,c}^s = S_c(S_f\left(\bar{x}_i\right)))=S(\bar{x}_i)$ and $p_{i,c}^t=T_c(T_f\left(x_i\right)))=T(x_i)$ are the outputs of student and teacher nets on class $c$, $y_{i,c}\in \{0,1\}$ is the corresponding arrhythmia ground truths. $S = \{S_f, S_c\}$ and $T = \{T_f, T_c\}$ are the student and the teacher nets with parameters $\theta_s = \{\theta_s^f, \theta_s^c\}$ and $\theta_t = \{\theta_t^f, \theta_t^c\}$. $T_f$ and $S_f$ are the feature extractors with parameters $\theta_t^f$ and $\theta_s^f$, while $T_c$ and $S_c$ are the classifiers with parameters $\theta_t^c$ and $\theta_s^c$. The input of the teacher net is a 12-lead ECG signals $x_i\in \mathcal{R}^{12\times L}$ sampled from the labeled dataset $\mathcal{D}_B = \{x_i,y_i\}_{i=1}^{N_B}$, where $L$ is the signal length and $y_i$ is the ground truth of $x_i$. $N_B$ is the sample size of the labeled dataset. The input of the student net is a 1-lead ECG signal $\bar{x}_i \in \mathcal{R}^{1\times L}$, which is generated by masking out the remaining leads of $x_i$ except lead I. In this study, the student net has much fewer parameters than the teacher net to ensure high inference efficiency. $\sigma(\cdot)$ is the sigmoid activation function and $\tau$ is a hyper-parameter that controls the distillation temperature.
\subsection{Region-aware knowledge distillation}
\label{sec:region}
Although traditional knowledge distillation can improve the arrhythmia detection accuracy of the student net,  its performance is still far inferior to that of the teacher net trained with 12-lead ECG signals. Here, we first summarize two important problems that resulted in the poor performance of previous methods and then propose the region-aware knowledge distillation technique to address them.  

\textbf{(1) 1-lead ECG signals cannot capture critical patterns that reveal the existence of certain arrhythmia, which limits the accuracy of the student net.} For example, left anterior fascicular block (LAnFB) cannot be reliably diagnosed using lead I alone, as its critical patterns, such as left axis deviation, are primarily observed in lead II and aVF  \cite{horwitz1975electrocardiographic,kashou2024electrical}. Consequently, it is difficult for the student net to recognize LAnFB from lead I ECG even with the assistance of the teacher net. Here, we address this problem using a simple but effective method. Specifically, we utilize a restoration net $R\left(\cdot\right)$ to recover the original 12-lead ECG signals from the 1-lead ECG signals. As shown in Section \ref{sec:pre}, it is pre-trained on a large-scale dataset and fine-tuned on the downstream dataset. Compared to raw 1-lead ECG $\bar{x}$, the restored 12-lead ECG signals $R\left(\bar{x}\right)$ contain more cardiac information for diagnosing a wide spectrum of arrhythmia, and is therefore more suitable for knowledge distillation. 

\textbf{(2) Previous knowledge distillation methods overlook the regional information within ECG signals, failing to emulate the diagnostic process of cardiologists.} The most important patterns of arrhythmia are abnormal waveforms and rhythms, which usually occur in specific regions (leads and timestamps) of ECG signals. These regions, which can be defined as regions of interest (ROIs), are critical for diagnosing specific arrhythmia. To fully exploit this property, we propose a region-aware knowledge distillation module. The underlying intuition is that distorting the ROIs should substantially affect the diagnosis results, whereas altering other regions should not. Therefore, ensuring prediction consistency between the student and teacher models on ECG signals with distorted ROIs offers an intuitive pipeline to improve distillation performance. However, accurately localizing arrhythmia-related ROIs across numerous ECG signals is challenging and often infeasible. As illustrated in Extended Data Fig.\ref{fig:flowchart_details}a, we provide a simple yet effective solution to realize the proposed region-aware knowledge distillation. We randomly replace a region of $x_i$ using the same region of $x_j$ to generate a new ECG signal $x_n$. Similarly, the restored signal $R(\bar{x}_i)$ can also be distorted by $R(\bar{x}_j)$ using the same pipeline. Note that the exchanged regions of $R(\bar{x}_i)$ and $x_i$ should be the same to allow for knowledge distillation. 
\begin{equation}
\label{Eq:mix}
x_n = M\odot x_i + (1 - M)\odot x_j,
\end{equation}
\begin{equation}
\label{Eq:mix_r}
R(\bar{x}_n) = M\odot R(\bar{x}_i) + (1 - M)\odot R(\bar{x}_j),
\end{equation}
where $M\in\{0,1\}^{12 \times L}$ is a binary mask matrix indicating the location and size of the exchanged region, and $\odot$ denotes the Hadamard product. Motivated by CutMix \cite{yun2019cutmix}, we generate the mask $M$ by sampling a bounding box coordinate $\mathcal{B} = \left(r_x, r_y, r_w, r_h\right)$ under a parameter $\lambda \sim Beta(\alpha, \alpha)$, where $\alpha$ is a shape parameter of the Beta distribution. Specifically, $r_x$ and $r_y$ are sampled uniformly from $\left[0, L\right]$ and $\left[0, 12\right]$, respectively. Then $r_w$ and $r_h$ can be computed by the formulas below,
\begin{equation}
\label{Eq:w}
r_w = L\sqrt{1-\lambda},
\end{equation}
\begin{equation}
\label{Eq:h}
r_h = 12\sqrt{1-\lambda},
\end{equation}
where parameter $\lambda$ controls the size of the exchanged region. We then minimize the divergence between the student's predictions $S(R(\bar{x}_n))$ and the teacher's predictions $T(x_n)$, which formulates the loss function of the proposed region-aware knowledge distillation,
\begin{equation}
\label{Eq:knowledge_region}
\mathcal{L}_{K}(\mathcal{D}_B,\theta_s,\theta_t)=-\frac{\tau^2}{NC}\sum^{N}_{n=1}\sum^{C}_{c=1}(1-\sigma(p_{n,c}^{t}/\tau))\log(1-\sigma(p_{n,c}^s/\tau))+\sigma(p_{n,c}^{t}/\tau)\log \sigma(p_{n,c}^{s}/\tau),
\end{equation}
\begin{equation}
\label{Eq:region_prediction}
p_{n}^s = S(R(\bar{x}_n))=S_c(S_f(R(\bar{x}_n))), p_{n}^t = T(x_n) =T_c(T_f(x_n)) .
\end{equation}

Regardless of whether the region $\mathcal{B}$ covers the ROIs for diagnosing arrhythmia from $x_i$ and $R(\bar{x}_i)$, the student will learn from the teacher's reactions to the distorted signal $x_n$. Specifically, if the ROIs critical for detecting a specific arrhythmia are removed, the teacher model will exhibit reduced confidence in that category. On the contrary, if new ROIs are added to the distorted signal, the teacher's confidence in the associated categories will increase. In all cases, the student will follow the teacher's reactions toward the distorted signals in order to minimize Eq.\ref{Eq:knowledge_region}. Utilizing this characteristic, the proposed region-aware distillation method can fully exploit the regional information for arrhythmia detection, boosting the performance of the student net. At the same time, we calculate the multi-label binary cross-entropy loss $\mathcal{L}_{Y}$ between the student net's predictions $R(\bar{x}_{n})$ and the ground truths $y_n$ as 
\begin{equation}
\label{Eq:y_mix}
y_n = y_i(1-\frac{r_w r_h}{12L})  + y_j\frac{r_w r_h}{12L},
\end{equation}
\begin{equation}
\label{Eq:supervised_loss}
\mathcal{L}_{Y}(\mathcal{D}_B,\theta_s)=-\frac{1}{NC}\sum^{N}_{n=1}\sum^{C}_{c=1}(1-y_{n,c})\log(1-\sigma(p_{n,c}^s))+y_{n,c}\log \sigma(p_{n,c}^{s}).
\end{equation}

\subsection{Improving distillation performance through cross-layer mutual information maximization}
\label{sec:mutual}
When the student and teacher model sizes differ significantly, the student's limited learning capacity prevents it from achieving robust performance \cite{huang2022knowledge}. Similar to many previous methods \cite{hinton2015distilling,park2019relational}, our region-aware knowledge distillation only matches the outputs of the student and teacher at specific layers, ignoring their reasoning pathways. This leads to an insufficient transfer of knowledge. Here, we propose a cross-layer mutual information maximization module to overcome this problem by strengthening the relationships between the student's reasoning and the teacher's diagnostic outputs $T\left(x\right)$. We use the intermediate features $S_f\left(R(\bar{x})\right)$ of the student net to represent its reasoning, and we want to make sure the student's reasoning process will lead to the teacher's output. Theoretically,  this objective can be achieved by maximizing the cross-layer mutual information $I(S_f\left(R(\bar{x})\right), T\left(x\right))$ between the student's intermediate features $S_f\left(R(\bar{x})\right)$ and the teacher's prediction $T\left(x\right)$, given as
\begin{equation}
\label{Eq:mutual}%\underset{\theta_s^f}
{\operatorname{max\thinspace}} I(Z_s, O_t) ={\operatorname{max\thinspace}}\iint p(o_t \mid z_s) p(z_s) \log \frac{p(o_t \mid z_s)}{p(o_t)} d z_s d o_t,
\end{equation}
where we use $Z_s$ to denote $S_f\left(R(\bar{x})\right)$ and $O_t$ to represent $T\left(x\right)$ for simplicity. Furthermore, it is easy to conclude that the mutual information defined in Eq.\ref{Eq:mutual} can be simplified as, 
%$\theta_s^f$ denotes the trainable parameters of the student feature extractor $S_f$
\begin{equation}
\label{Eq:mutual_sim}
I(Z_s, O_t) = KL(p(o_t\mid z_s)p(z_s)\mid\mid p(o_t)p(z_s)),
\end{equation}
where $KL(\cdot\mid\mid\cdot)$ denotes the Kullback-Leibler divergence between two probabilistic distributions. Consequently, the objective in Eq.\ref{Eq:mutual} can be converted to maximize the Kullback-Leibler divergence between $p(o_t\mid z_s)p(z_s)$ and $p(o_t)p(z_s)$ as  
\begin{equation}
\label{Eq:mutual_kl}
{\operatorname{max\thinspace}}KL(p(o_t\mid z_s)p(z_s)\mid\mid p(o_t)p(z_s)). 
\end{equation}
However, maximizing the Kullback-Leibler divergence might lead to an unstable optimization trajectory due to its unlimited upper bound. Fortunately, the Jensen–Shannon divergence is a symmetrized and smoothed version of the Kullback-Leibler divergence, providing a better option to approximate the mutual information, which can be defined as
\begin{equation}
\label{Eq:JS}
JS(P, Q)=\frac{1}{2} K L\left(P \| \frac{P+Q}{2}\right)+\frac{1}{2} K L\left(Q \| \frac{P+Q}{2}\right).
\end{equation}
Consequently, we can rewrite Eq.\ref{Eq:mutual_kl} as,
\begin{equation}
\label{Eq:mutual_js}
{\operatorname{max\thinspace}}JS(p(o_t\mid z_s)p(z_s)\mid\mid p(o_t)p(z_s)). 
\end{equation}
Motivated by \cite{nowozin2016f, hjelm2018learning}, we can introduce a discriminator $D$ with parameter $\theta_d$ to estimate the Jensen–Shannon divergence, and maximize it using the student feature extractor $S_f$,
\begin{equation}
\label{Eq:mutual_final}
{\operatorname{max\thinspace}}\mathbb{E}_{(o_t, z_s)  \sim p(o_t\mid z_s)p(z_s)}[\log \sigma(D(o_t, z_s))]+\mathbb{E}_{(o_t, z_s) \sim p(o_t)p(z_s)}[\log (1-\sigma(D(o_t, z_s))],
\end{equation}
where $\theta_d$ denotes the parameter of the discriminator and $\sigma$ is the sigmoid activation. Specifically, the discriminator should output a positive prediction for a paired sample $(z_s^i, o_t^i)$ while providing a negative prediction for an unpaired sample $(z_s^i, o_t^j), i\neq j$, where $z_s^i = S_f(R(\bar{x}^i)), o_t^i = T(x^i), o_t^j = T(x^j), i\neq j$. During this process, the cross-layer mutual information $I(Z_s, O_t)$ is estimated. At the same time, student feature extractor $S_f$ should adjust the extracted feature distribution to maximize it. For mini-batch training, Eq.\ref{Eq:mutual_final} can be rewritten as minimizing a discriminator loss,
\begin{equation}
\label{Eq:mutual_final_batch}
{\operatorname{min\thinspace}}\mathcal{L}_D(\mathcal{D}_B,\theta_s,\theta_d)={\operatorname{min\thinspace}} -\frac{1}{N} \sum_{n=1}^{N}\left[\log(\sigma(D(z_s^n, o_t^n))) + \log(1-\sigma(D(z_s^n, o_t^{N-n})))\right]
\end{equation}

\subsection{Incorporating unlabeled samples for semi-supervised knowledge distillation}
Collecting sufficient well-labeled ECG data is time-consuming and expensive, which results in the label scarcity problem in clinical practice. Without sufficient labeled samples, it will be difficult for the student net to recover the performance of the teacher net during the knowledge distillation process \cite{phuong2019towards}. Fortunately, unlabeled samples are much easier to collect compared with the labeled samples, which could provide sufficient information to enhance the student's performance \cite{zhou2023semi}. However, previous methods only use labeled samples for knowledge transfer \cite{hinton2015distilling,park2019relational,zhao2022decoupled,qin2023mvkt}, which limits their robustness and flexibility in clinical applications. Here, we incorporate unlabeled samples for knowledge distillation and implement our framework in a semi-supervised manner. First, we compute the region-aware knowledge distillation loss $\mathcal{L}_K(\mathcal{D}_B, \theta_s, \theta_t)$, the discriminator loss $\mathcal{L}_D(\mathcal{D}_B, \theta_s, \theta_d)$ and the supervised multi-label classification loss $\mathcal{L}_Y(\mathcal{D}_B, \theta_s)$ on the labeled training set $\mathcal{D}_B = \{x_i,y_i\}_{i=1}^{N_B}$. 
\begin{equation}
\label{Eq:labeled}
\mathcal{L}^{\text{labeled}} = \mathcal{L}_Y(\mathcal{D}_B, \theta_s) + \alpha \mathcal{L}_K(\mathcal{D}_B, \theta_s, \theta_t) + \beta \mathcal{L}_D(\mathcal{D}_B, \theta_s, \theta_d), 
\end{equation}
where $\alpha$ and $\beta$ are two hyper-parameters that control the importance of different losses. Considering that label information is absent in the unlabeled dataset $\mathcal{D}_U = \{x_i, -\}_{i=1}^{N_U}$,  we only calculate the knowledge distillation loss and the discriminator loss,  
\begin{equation}
\label{Eq:unlabeled}
\mathcal{L}^{\text{unlabeled}} = \alpha \mathcal{L}_K(\mathcal{D}_U, \theta_s, \theta_t) + \beta \mathcal{L}_D(\mathcal{D}_U, \theta_s, \theta_d).
\end{equation}
Combining Eq.\ref{Eq:labeled} and Eq.\ref{Eq:unlabeled}, we can formulate the overall loss function of the proposed framework, given as 
\begin{equation}
\begin{split}
\label{Eq:final_objective}
\mathcal{L} &= \mathcal{L}^{\text{labeled}} + \mathcal{L}^{\text{unlabeled}}\\&=
\mathcal{L}_Y(\mathcal{D}_B, \theta_s) + \alpha \left[\mathcal{L}_K(\mathcal{D}_B, \theta_s, \theta_t) + \mathcal{L}_K(\mathcal{D}_U, \theta_s, \theta_t)\right] \\&+ \beta \left[\mathcal{L}_D(\mathcal{D}_B, \theta_s, \theta_d)+\mathcal{L}_D(\mathcal{D}_U, \theta_s, \theta_d)\right].
\end{split}
\end{equation}
During the end-to-end training process, $\theta_s$ and $\theta_d$ are updated to minimize $\mathcal{L}$ while $\theta_t$ is fixed. 

\subsection{Implementation details}
\label{sec:pre}
The restoration and CVDs diagnostic models are pre-trained on a public ECG dataset with 2,322,513 12-lead ECG signals from 1,558,772 patients \cite{ribeiro2019tele}. A held-out dataset with 827 ECG signals from 827 patients is adopted as the validation set \cite{ ribeiro2020automatic}. The duration of the ECG signals ranges from 7 to 10 seconds, and their sampling rate ranges from 300 to 600 Hz. We formulate the restoration net using a UNet-based architecture \cite{ronneberger2015u,yoon2024classification, chen2024multi} with 0.36 to 5.71 million trainable parameters. The detailed configuration of the restoration model is presented in the Supplementary Materials. As shown in Extended Data Fig.\ref{fig:flowchart_details}d, during the pre-training process, we convert the original 12-lead ECG signals into 1-lead ECG signals by masking out the remaining leads except lead I. For a given 1-lead ECG signal, the restoration net is used to restore the corresponding 12-lead ECG signals. Its parameters are updated to minimize the mean squared error between the generated 12-lead ECG signals and the original 12-lead ECG signals. The pre-training process is terminated when the restoration error on the validation set does not decrease for 10 epochs. The large-scale teacher net CVDs detection is also pre-trained on the same dataset using 12-lead ECG signals and the corresponding multi-label ground truths. In this study, we use the Convolution-Transformer network proposed in \cite{zhou2024computation} to formulate the architecture of the teacher and student nets. Specifically, the teacher net $T$ consists of a feature extraction module $T_f$ and a classification module $T_c$. The student net $S = \{S_f, S_c\}$ shares the same architecture with the teacher net but has much fewer parameters to ensure high inference speed on low-level devices. To be specific, the teacher net and student net have 50.5 million and 0.26 to 1.60 million trainable parameters, respectively. AdamW optimizer \cite{loshchilov2017decoupled} is used for the pre-training process with a learning rate of 1e-3 and a batch size of 1024. The detailed configurations are presented in the Supplementary Materials. Their pre-training processes are terminated when the classification error on the validation set does not decrease for 10 epochs.

On the downstream datasets, the aforementioned restoration net and CVDs diagnostic models are fine-tuned using the same loss functions defined in the pre-training process. Specifically, the restoration net is fine-tuned using all the ECG signals from the downstream dataset, while the teacher net is only fine-tuned with the labeled set (Extended Data Fig.\ref{fig:flowchart_details}e). Subsequently, they are used for generating low-cost cardiac intelligence with the LiteHeart framework. AdamW optimizer \cite{loshchilov2017decoupled} is used for the fine-tuning process with a learning rate of 2e-3 and a batch size of 128. Finally, their best checkpoints on the validation set are selected for the following knowledge distillation process. Using the labeled dataset $D_B$ and the unlabeled dataset $D_U$, the proposed LiteHeart framework generates a small-scale student net, which can formulate low-cost cardiac intelligence together with the restoration net. During this process, the student net is optimized using the AdamW optimizer with a learning rate of 2e-3 and a batch size of 128, while the teacher net and the restoration net are fixed.

\section{Data availability}
The wearable ECG dataset used in our study is publicly available at \href{https://www.scidb.cn/detail?dataSetId=58c4a92d5a01414390a78160d335380d}{https://www.scidb.cn/detail?dataSetId=58c4a92d5a01414390a78160d335380d}. Restrictions apply to the availability of the large-scale dataset set for model pertaining and requests to access it must be submitted and reviewed individually by the Telehealth Network of Minas Gerais (antonio.ribeiro@ebserh.gov.br) for academic use only. The Georgia 12-lead ECG signals Challenge (G12EC) database can be accessed at \href{https://www.kaggle.com/datasets/physionet/georgia-12lead-ecg-challenge-database}{https://www.kaggle.com/datasets/physionet/georgia-12lead-ecg-challenge-database}. The Chapman-Shaoxing database, the Ningbo database, and the Physikalisch-Technische Bundesanstalt (PTB-XL) database can be freely downloaded from \href{https://figshare.com/collections/ChapmanECG/4560497/2}{https://figshare.com/collections/ChapmanECG/4560497/2} , \href{https://physionet.org/content/ecg-arrhythmia/1.0.0/}{https://physionet.org/content/ecg-arrhythmia/1.0.0/} , \href{https://physionet.org/content/ptb-xl/1.0.3/}{https://physionet.org/content/ptb-xl/1.0.3/}.

\section{Code availability}
All relevant models were implemented based on Python language and a popular deep-learning framework Pytorch. Our code will be released at \href{https://github.com/KAZABANA/LiteHeart}{https://github.com/KAZABANA/LiteHeart} after publication.





\bibliography{sn-bibliography}% common bib file

\bmhead{Acknowledgments}
This work was funded in part by the InnoHK initiative of the Innovation and Technology Commission of the Hong Kong Special Administrative Region Government, in part by the National Natural Science Foundation of China (22322816), and the City University of Hong Kong Project (9610640).

\bmhead{Author contributions}
R.Z. and Y.D. conceived and designed the study. R.Z. wrote the code and conducted the experiments. R.Z. and Y.D. co-wrote the manuscript. J.D. and Y.Z. helped to revise the manuscript. Y.D., J.D., and Y.Z. were senior advisors to the project. All authors contributed to the results interpretation and final manuscript preparation.

\bmhead{Competing interests}
The authors declare that they have no conflicts of interest.

\begin{extendfigure*}[p]
\begin{center}
\includegraphics[width=1\textwidth]{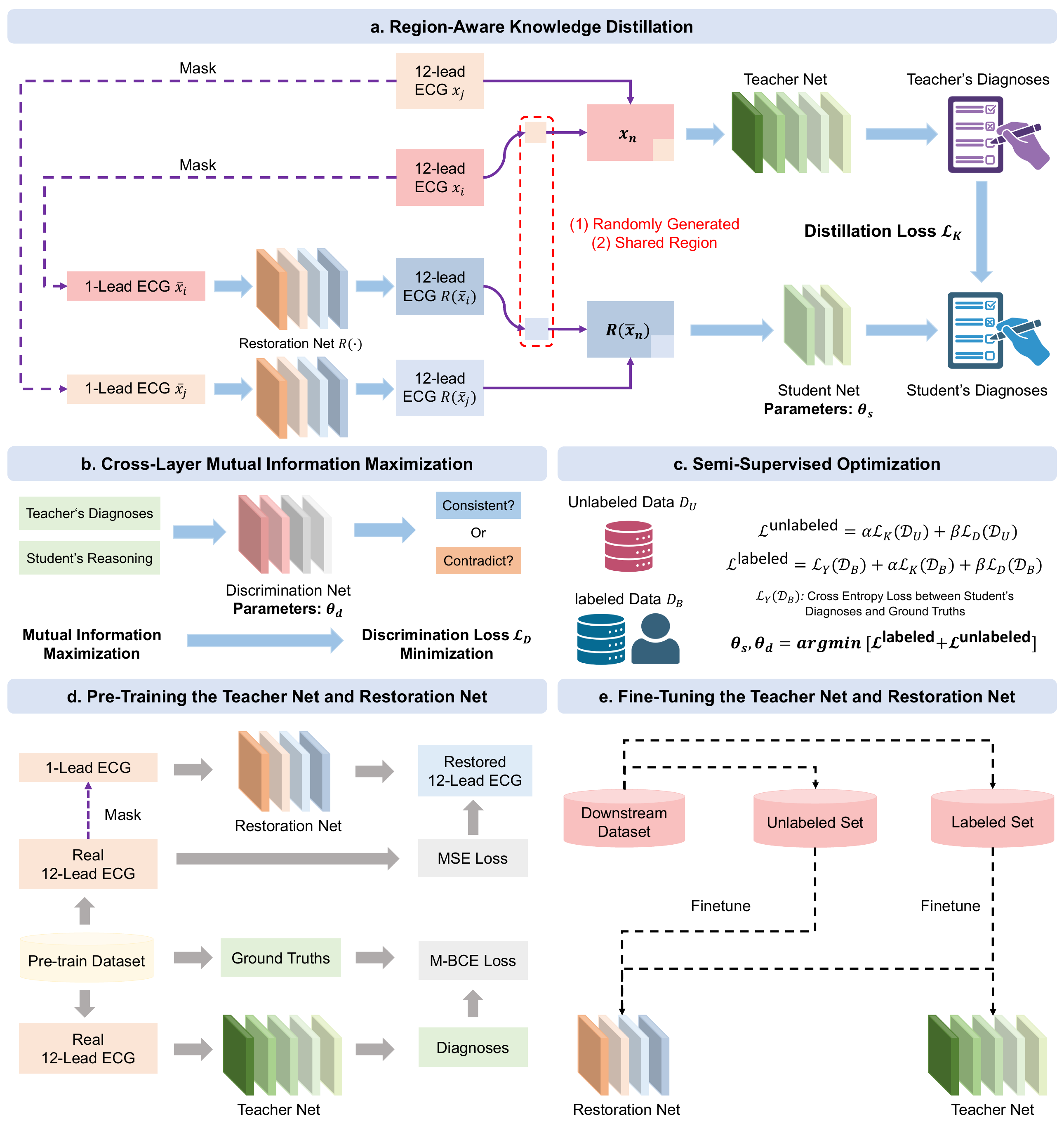}
\end{center}
\caption{Overview of LiteHeart.}
\label{fig:flowchart_details}
\end{extendfigure*}

\begin{extendtable*}[p]
\setlength{\tabcolsep}{0.15em}
\fontsize{8}{9.5}\selectfont
\caption{Description of the cardiovascular diseases analyzed in our study.}
\label{tab:anotation}
\color{black}
\begin{tabular*}{\hsize}{@{}lclc@{}}
\toprule
\textbf{Original annotation} & \textbf{Abbreviations} &
\textbf{Original annotation} & \textbf{Abbreviations}\\
\midrule
\multicolumn{4}{c}{\textbf{SMU Dataset}}\\
\midrule
normal ECG&NECG&sinus rhythm&SR\\
Q wave abnormal&QAb&q wave abnormal&qAb\\
poor R wave progression&PRP&ST elevation&STE\\
ST depression&STD&T wave abnormal&TAb\\
sinus tachycardia&ST&sinus bradycardia&SB\\
atrial fibrillation&AF&atrial flutter&AFL\\
premature ventricular contractions&PVC&first-degree atrioventricular block&IAVB\\
bundlebranchblock&BBB&complete right bundle branch block&CRBBB\\
 incomplete right bundle branch block&IRBBB&partial right bundle branch block&PRBBB\\
heart enlargement and hypertrophy&HEH&high left ventricular voltage&HLVV\\
paced rhythm&PR&premature atrial contractions&PAC\\sinus arrhythmia&SA&&\\
\midrule
\multicolumn{4}{c}{\textbf{G12EC Dataset}}\\
\midrule
atrial fibrillation&AF&1st degree av block&IAVB\\
incomplete right bundle branch block&IRBBB&left axis deviation&LAD\\
left bundle branch block&LBBB&low qrs voltages&LQRSV\\
nonspecific intraventricular conduction disorder&NSIVCB&sinus rhythm&NSR\\
premature atrial contraction&PAC&prolonged qt interval&LQT\\
qwave abnormal&QAb&right bundle branch block&RBBB\\
sinus arrhythmia&SA&sinus bradycardia&SB\\
sinus tachycardia&STach&t wave abnormal&TAb\\
t wave inversion&TInv&ventricular premature beats&VPB\\
\midrule
\multicolumn{4}{c}{\textbf{PTB-XL Dataset}}\\
\midrule
atrial fibrillation&AF&complete right bundle branch block&CRBBB\\
1st degree av block&IAVB&incomplete right bundle branch block&IRBBB\\
left axis deviation&LAD&left anterior fascicular block&LAnFB\\
left bundle branch block&LBBB&nonspecific intraventricular conduction disorder&NSIVCB\\
sinus rhythm&NSR&premature atrial contraction&PAC\\
pacing rhythm&PR&prolonged pr interval&LPR\\
qwave abnormal&QAb&right axis deviation&RAD\\
sinus arrhythmia&SA&sinus bradycardia&SB\\
sinus tachycardia&STach&t wave abnormal&TAb\\
t wave inversion&TInv&&\\
\midrule
\multicolumn{4}{c}{\textbf{Ningbo Dataset}}\\
\midrule
atrial flutter&AFL&bundle branch block&BBB\\
complete left bundle branch block&CLBBB&complete right bundle branch block&CRBBB\\
1st degree av block&IAVB&incomplete right bundle branch block&IRBBB\\
left axis deviation&LAD&left anterior fascicular block&LAnFB\\
low qrs voltages&LQRSV&nonspecific intraventricular conduction disorder&NSIVCB\\
sinus rhythm&NSR&premature atrial contraction&PAC\\
pacing rhythm&PR&poor R wave Progression&PRWP\\
premature ventricular contractions&PVC&prolonged qt interval&LQT\\
qwave abnormal&QAb&right axis deviation&RAD\\
sinus arrhythmia&SA&sinus bradycardia&SB\\
sinus tachycardia&STach&t wave abnormal&TAb\\
t wave inversion&TInv&&\\
\midrule
\multicolumn{4}{c}{\textbf{Chapman Dataset}}\\
\midrule
atrial fibrillation&AF&atrial flutter&AFL\\
1st degree av block&IAVB&left axis deviation&LAD\\
left bundle branch block&LBBB&low qrs voltages&LQRSV\\
nonspecific intraventricular conduction disorder&NSIVCB&sinus rhythm&NSR\\
premature atrial contraction&PAC&qwave abnormal&QAb\\
right axis deviation&RAD&right bundle branch block&RBBB\\
sinus bradycardia&SB&sinus tachycardia&STach\\
t wave abnormal&TAb&ventricular premature beats&VPB\\
\bottomrule
\end{tabular*}
\end{extendtable*}

\begin{extendfigure*}[p]
\begin{center}
\includegraphics[width=1\textwidth]{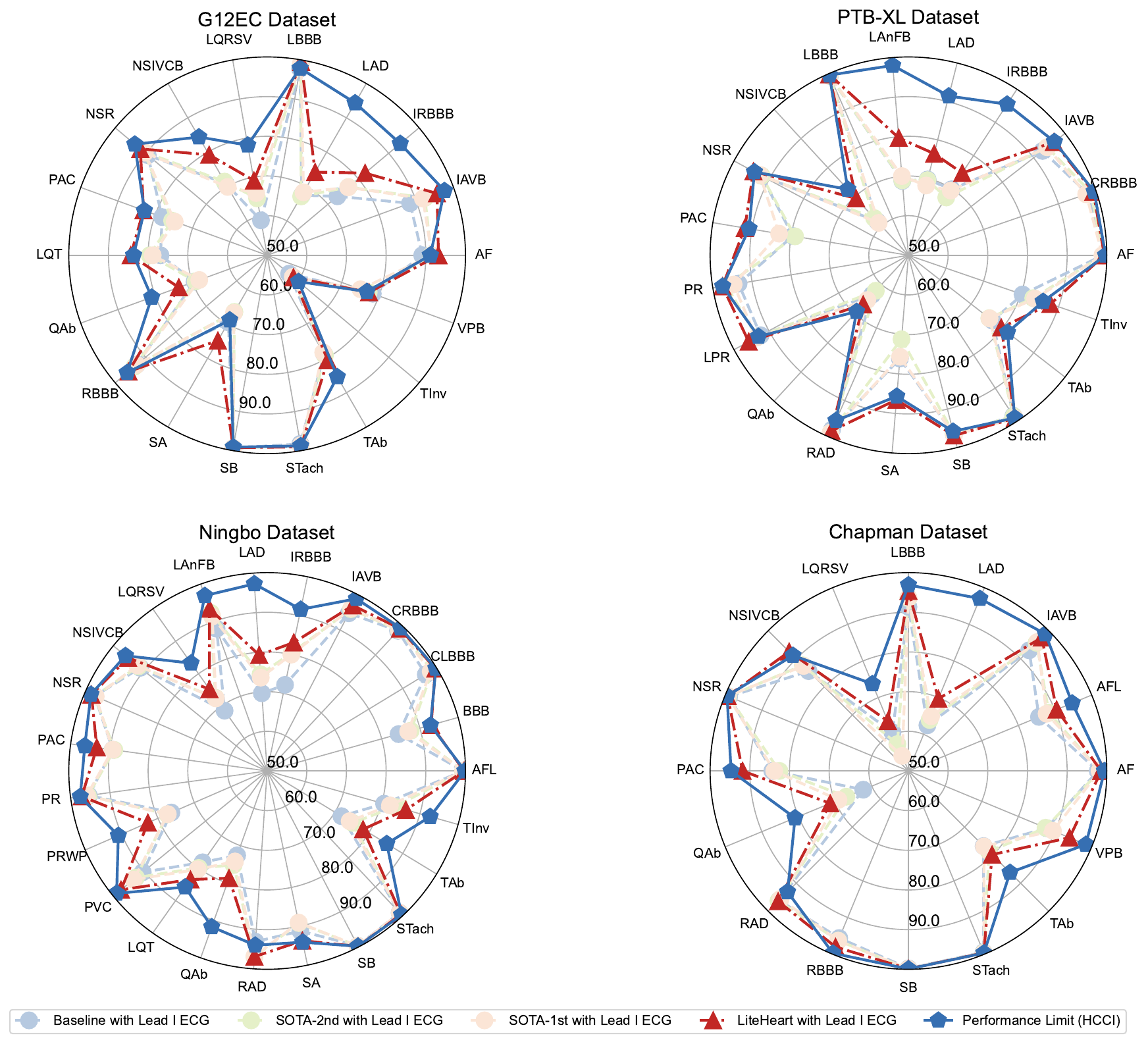}
\end{center}
\caption{AUC of different models on various CVDs from the G12EC, PTB, Ningbo, Chapman datasets.}
\label{fig:radar_extend}
\end{extendfigure*}

\begin{extendfigure*}[p]
\begin{center}
\includegraphics[width=1\textwidth]{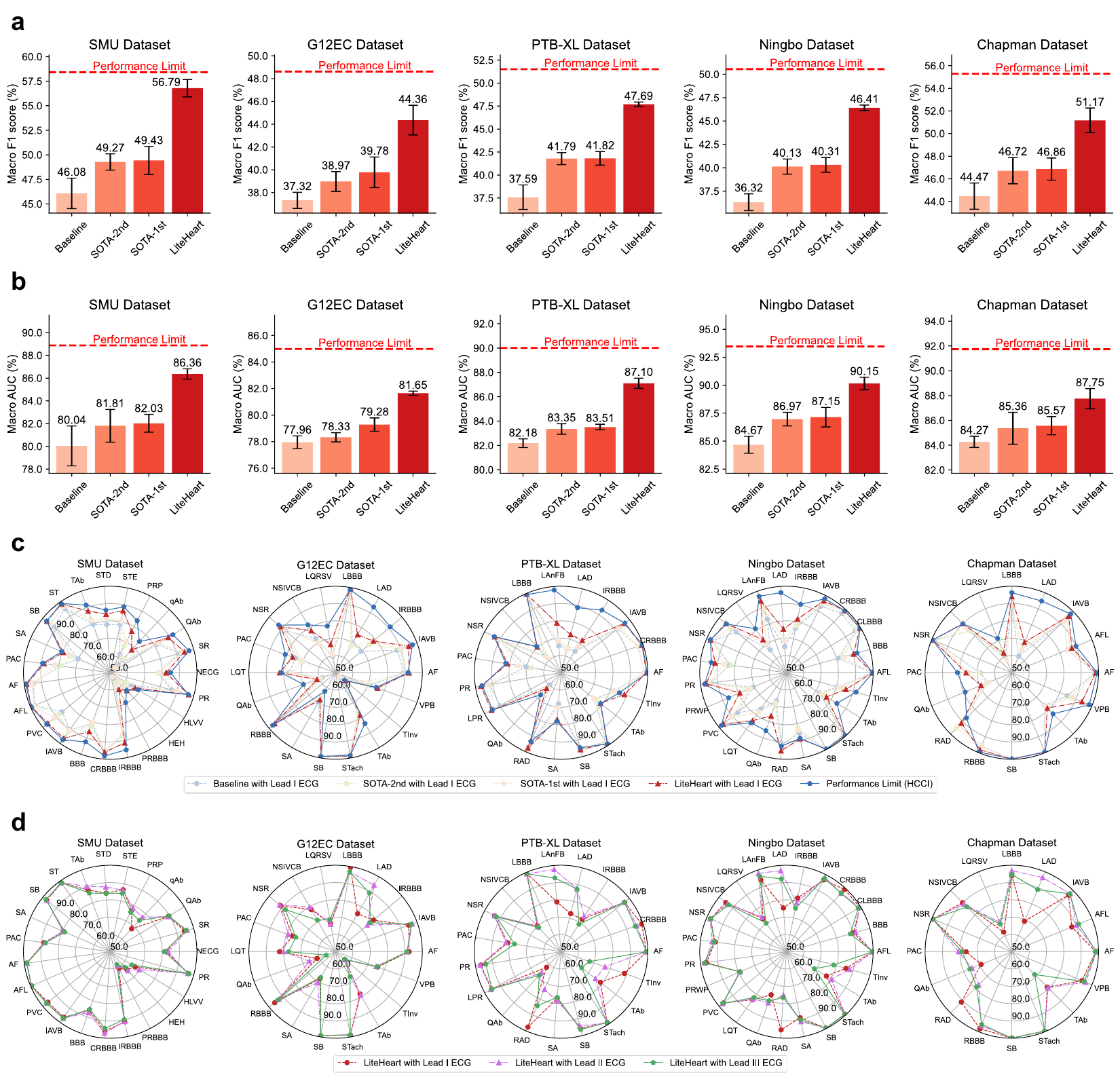}
\end{center}
\caption{\textbf{a-b.} Macro F1 score and macro AUC of the low-cost cardiac intelligence generated by the proposed LiteHeart and the SOTA methods on different downstream datasets. \textbf{c.} AUC of different models on various CVDs. \textbf{d.} The impact of ECG lead selection on the diagnostic performance of the low-cost cardiac intelligence. Note that the ratio of the labeled samples for model training is set to 5\% in these experiments.}
\label{fig:compare_tiny_005}
\end{extendfigure*}

\begin{extendfigure*}[p]
\begin{center}
\includegraphics[width=1\textwidth]{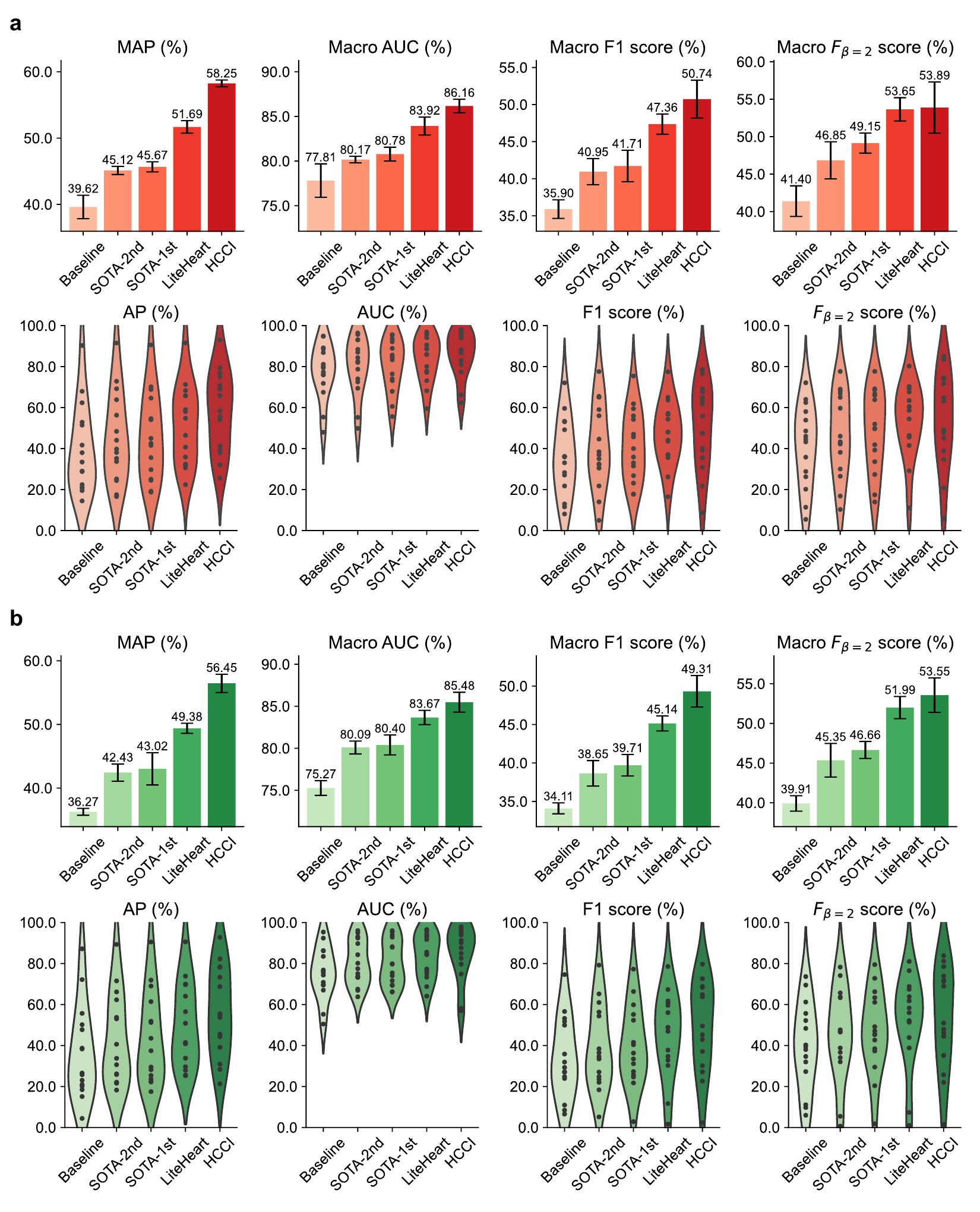}
\end{center}
\caption{Cross-device external validation of different models under very limited supervision. \textbf{a.} The diagnostic performance of different models with 5\% labeled samples. \textbf{b.} The diagnostic performance of different models with 2.5\% labeled samples.}
\label{fig:external_extend}
\end{extendfigure*}

\begin{extendtable*}[p]
\fontsize{7}{11}\selectfont
\setlength{\tabcolsep}{0.1em}
\begin{center}
\caption{Comparison results between LiteHeart and the SOTA methods with 10\% labeled samples. The mean performance and standard deviations on 5 databases are shown across 4 seeds.}
\label{tab:compare}
\scalebox{1}{
\color{black}
\begin{tabular*}{\hsize}{@{}@{\extracolsep{\fill}}lccccccccccc@{}}
\toprule
Methods & KD & DKD & CC & RKD & DIST & PKT & PSM & WKD  & MVKT & \textbf{LiteHeart} \\
\midrule
\multicolumn{11}{c}{\textbf{Ranking loss} (The smaller, the better)}\\
\midrule
SMU & 0.09$\pm$0.01&0.09$\pm$0.01&0.09$\pm$0.01&0.09$\pm$0.01&0.09$\pm$0.00&0.09$\pm$0.00&0.09$\pm$0.01&0.09$\pm$0.00&0.09$\pm$0.00&\textbf{0.06$\pm$0.00}\\
G12EC & 0.13$\pm$0.00&0.13$\pm$0.01&0.13$\pm$0.01&0.13$\pm$0.01&0.13$\pm$0.01&0.13$\pm$0.00&0.13$\pm$0.00&0.12$\pm$0.00&0.12$\pm$0.01&\textbf{0.10$\pm$0.00}\\
PTB-XL & 0.05$\pm$0.00&0.05$\pm$0.00&0.05$\pm$0.00&0.05$\pm$0.00&0.05$\pm$0.00&0.05$\pm$0.00&0.05$\pm$0.00&0.05$\pm$0.00&0.04$\pm$0.00&\textbf{0.03$\pm$0.00}\\
Ningbo & 0.04$\pm$0.00&0.04$\pm$0.00&0.04$\pm$0.00&0.04$\pm$0.00&0.04$\pm$0.00&0.04$\pm$0.00&0.04$\pm$0.00&0.04$\pm$0.00&0.04$\pm$0.00&\textbf{0.03$\pm$0.00}\\
Chapman & 0.05$\pm$0.00&0.06$\pm$0.00&0.05$\pm$0.00&0.05$\pm$0.00&0.06$\pm$0.00&0.08$\pm$0.00&0.05$\pm$0.00&0.05$\pm$0.00&0.05$\pm$0.00&\textbf{0.04$\pm$0.00}\\
\midrule
\multicolumn{11}{c}{\textbf{Coverage} (The smaller, the better)}\\
\midrule
SMU & 7.29$\pm$0.22&7.36$\pm$0.27&7.28$\pm$0.19&7.48$\pm$0.22&7.31$\pm$0.15&7.23$\pm$0.15&7.34$\pm$0.14&7.19$\pm$0.24&7.12$\pm$0.19&\textbf{6.23$\pm$0.07}\\
G12EC & 4.57$\pm$0.06&4.67$\pm$0.14&4.60$\pm$0.08&4.64$\pm$0.13&4.73$\pm$0.16&4.60$\pm$0.09&4.60$\pm$0.13&4.53$\pm$0.06&4.55$\pm$0.16&\textbf{3.96$\pm$0.10}\\
PTB-XL & 3.22$\pm$0.08&3.20$\pm$0.10&3.19$\pm$0.07&3.29$\pm$0.07&3.14$\pm$0.07&3.23$\pm$0.08&3.20$\pm$0.07&3.11$\pm$0.07&3.08$\pm$0.02&\textbf{2.74$\pm$0.02}\\
Ningbo & 3.24$\pm$0.12&3.21$\pm$0.11&3.17$\pm$0.09&3.24$\pm$0.05&3.29$\pm$0.09&3.20$\pm$0.12&3.18$\pm$0.09&3.19$\pm$0.10&3.22$\pm$0.07&\textbf{2.71$\pm$0.04}\\
Chapman & 2.72$\pm$0.05&2.84$\pm$0.07&2.76$\pm$0.07&2.79$\pm$0.07&2.84$\pm$0.05&3.23$\pm$0.08&2.79$\pm$0.09&2.70$\pm$0.06&2.72$\pm$0.06&\textbf{2.42$\pm$0.01}\\
\midrule
\multicolumn{11}{c}{\textbf{MAP} (The greater, the better)}\\
\midrule
SMU & 56.31$\pm$1.56&56.60$\pm$1.77&56.44$\pm$1.53&55.85$\pm$1.34&56.62$\pm$1.13&56.77$\pm$1.16&56.79$\pm$1.25&57.33$\pm$1.08&57.60$\pm$0.95&\textbf{64.86$\pm$0.71}\\
G12EC & 43.32$\pm$1.01&42.74$\pm$1.18&42.38$\pm$1.14&42.31$\pm$0.80&42.30$\pm$1.19&42.68$\pm$1.25&42.53$\pm$0.91&42.94$\pm$0.43&42.94$\pm$1.08&\textbf{48.65$\pm$1.49}\\
PTB-XL & 44.70$\pm$0.75&44.60$\pm$0.45&44.49$\pm$0.79&44.15$\pm$0.78&44.99$\pm$0.56&45.05$\pm$0.86&45.01$\pm$0.41&45.02$\pm$0.47&45.29$\pm$0.57&\textbf{51.09$\pm$0.27}\\
Ningbo & 43.60$\pm$1.11&43.00$\pm$0.64&43.61$\pm$0.56&43.02$\pm$1.13&44.29$\pm$0.81&43.47$\pm$0.55&43.43$\pm$0.63&44.09$\pm$0.85&44.08$\pm$0.95&\textbf{50.90$\pm$0.63}\\
Chapman & 51.16$\pm$0.76&50.10$\pm$0.86&49.60$\pm$0.25&49.42$\pm$1.28&49.53$\pm$0.94&39.44$\pm$1.70&50.30$\pm$0.86&51.44$\pm$0.97&50.99$\pm$0.89&\textbf{55.08$\pm$0.51}\\
\midrule
\multicolumn{11}{c}{\textbf{Macro AUC} (The greater, the better)}\\
\midrule
SMU & 83.48$\pm$1.35&83.57$\pm$1.25&83.60$\pm$1.11&82.71$\pm$0.94&83.14$\pm$0.78&83.72$\pm$0.62&83.32$\pm$0.81&84.20$\pm$0.41&83.86$\pm$0.74&\textbf{87.70$\pm$0.54}\\
G12EC & 80.61$\pm$0.44&80.69$\pm$0.85&80.50$\pm$0.39&79.76$\pm$0.55&80.37$\pm$0.78&80.36$\pm$0.62&80.59$\pm$0.70&80.28$\pm$0.51&80.48$\pm$0.77&\textbf{83.96$\pm$0.73}\\
PTB-XL & 84.27$\pm$0.47&84.37$\pm$0.77&84.60$\pm$0.51&83.55$\pm$0.70&84.36$\pm$0.57&84.27$\pm$0.77&84.60$\pm$0.14&85.17$\pm$0.36&85.66$\pm$0.43&\textbf{88.95$\pm$0.20}\\
Ningbo & 88.20$\pm$1.21&88.63$\pm$0.79&88.62$\pm$0.82&87.28$\pm$0.42&88.58$\pm$0.85&88.75$\pm$0.61&88.80$\pm$0.86&88.59$\pm$0.46&88.69$\pm$0.76&\textbf{91.31$\pm$0.46}\\
Chapman & 87.14$\pm$0.78&86.43$\pm$0.45&87.02$\pm$0.94&86.23$\pm$0.90&86.89$\pm$1.22&82.57$\pm$1.67&86.97$\pm$0.40&87.01$\pm$0.79&87.43$\pm$0.42&\textbf{89.88$\pm$0.40}\\
\midrule
\multicolumn{11}{c}{\textbf{Macro F1 score} (The greater, the better)}\\
\midrule
SMU & 50.55$\pm$1.33&51.01$\pm$0.52&50.89$\pm$0.75&50.56$\pm$1.11&51.66$\pm$0.52&51.50$\pm$0.90&50.98$\pm$0.74&52.10$\pm$0.56&52.09$\pm$1.92&\textbf{59.20$\pm$0.47}\\
G12EC & 41.78$\pm$0.73&42.32$\pm$1.60&41.79$\pm$0.67&41.08$\pm$1.01&41.42$\pm$1.36&41.11$\pm$1.44&41.84$\pm$1.16&41.41$\pm$0.53&42.58$\pm$1.48&\textbf{46.95$\pm$1.11}\\
PTB-XL & 43.83$\pm$0.70&43.54$\pm$0.49&43.61$\pm$0.96&43.15$\pm$0.56&44.02$\pm$0.83&43.91$\pm$0.90&44.02$\pm$0.36&44.51$\pm$0.47&43.76$\pm$0.64&\textbf{49.95$\pm$0.37}\\
Ningbo & 42.61$\pm$0.84&42.69$\pm$0.34&42.95$\pm$0.44&41.88$\pm$0.16&43.16$\pm$0.65&43.12$\pm$0.45&43.18$\pm$0.40&43.13$\pm$0.35&43.38$\pm$0.61&\textbf{49.51$\pm$0.66}\\
Chapman & 48.95$\pm$1.12&49.00$\pm$1.23&48.18$\pm$1.26&48.61$\pm$1.23&49.21$\pm$0.56&39.85$\pm$1.34&49.05$\pm$1.10&48.79$\pm$0.70&49.66$\pm$0.66&\textbf{53.93$\pm$0.47}\\
\midrule
\multicolumn{11}{c}{\textbf{Macro $F_{\beta=2}$ score} (The greater, the better)}\\
\midrule
SMU & 60.32$\pm$1.77&60.61$\pm$1.01&60.20$\pm$1.22&59.92$\pm$1.24&60.38$\pm$1.23&60.65$\pm$1.48&60.62$\pm$1.00&60.61$\pm$1.38&61.53$\pm$0.94&\textbf{66.93$\pm$0.53}\\
G12EC & 52.06$\pm$0.82&52.85$\pm$1.21&52.31$\pm$0.84&51.64$\pm$0.68&51.66$\pm$0.86&51.72$\pm$0.99&52.28$\pm$0.81&51.64$\pm$0.45&52.04$\pm$1.02&\textbf{56.58$\pm$0.85}\\
PTB-XL & 52.46$\pm$0.46&52.52$\pm$0.52&52.66$\pm$0.89&52.41$\pm$0.74&52.76$\pm$0.99&52.83$\pm$0.78&52.74$\pm$0.21&53.44$\pm$0.64&52.84$\pm$0.56&\textbf{58.70$\pm$0.40}\\
Ningbo & 51.34$\pm$1.16&51.49$\pm$0.50&51.65$\pm$1.18&50.75$\pm$0.32&51.80$\pm$0.73&52.25$\pm$0.43&51.55$\pm$0.85&52.51$\pm$0.45&52.27$\pm$0.43&\textbf{58.06$\pm$0.43}\\
Chapman & 56.94$\pm$0.83&57.52$\pm$1.26&56.02$\pm$1.42&56.64$\pm$0.80&56.92$\pm$0.81&48.41$\pm$1.44&55.75$\pm$1.48&57.43$\pm$0.86&57.53$\pm$0.78&\textbf{61.86$\pm$0.95}\\
\bottomrule
\end{tabular*}
}
\end{center}
\end{extendtable*}

\begin{extendtable*}[p]
\fontsize{7}{11}\selectfont
\setlength{\tabcolsep}{0.1em}
\begin{center}
\caption{Comparison results between LiteHeart and the SOTA methods with 5\% labeled samples. The mean performance and standard deviations on 5 databases are shown across 4 seeds.}
\label{tab:compare_005}
\scalebox{1}{
\color{black}
\begin{tabular*}{\hsize}{@{}@{\extracolsep{\fill}}lccccccccccc@{}}
\toprule
Methods & KD & DKD & CC & RKD & DIST & PKT & PSM & WKD  & MVKT & \textbf{LiteHeart} \\
\midrule
\multicolumn{11}{c}{\textbf{Ranking loss} (The smaller, the better)}\\
\midrule
SMU & 0.10$\pm$0.01&0.10$\pm$0.01&0.10$\pm$0.01&0.10$\pm$0.01&0.10$\pm$0.00&0.10$\pm$0.01&0.10$\pm$0.01&0.10$\pm$0.01&0.10$\pm$0.00&\textbf{0.07$\pm$0.00}\\
G12EC & 0.14$\pm$0.01&0.14$\pm$0.00&0.14$\pm$0.00&0.14$\pm$0.01&0.14$\pm$0.01&0.14$\pm$0.01&0.14$\pm$0.01&0.14$\pm$0.00&0.13$\pm$0.00&\textbf{0.11$\pm$0.00}\\
PTB-XL & 0.05$\pm$0.00&0.05$\pm$0.00&0.05$\pm$0.00&0.05$\pm$0.00&0.05$\pm$0.00&0.05$\pm$0.00&0.05$\pm$0.00&0.05$\pm$0.00&0.05$\pm$0.00&\textbf{0.04$\pm$0.00}\\
Ningbo & 0.05$\pm$0.00&0.05$\pm$0.00&0.05$\pm$0.00&0.05$\pm$0.00&0.05$\pm$0.00&0.05$\pm$0.00&0.05$\pm$0.00&0.05$\pm$0.00&0.05$\pm$0.00&\textbf{0.04$\pm$0.00}\\
Chapman & 0.06$\pm$0.00&0.06$\pm$0.00&0.06$\pm$0.00&0.06$\pm$0.00&0.06$\pm$0.01&0.06$\pm$0.01&0.06$\pm$0.00&0.06$\pm$0.01&0.06$\pm$0.00&\textbf{0.05$\pm$0.00}\\
\midrule
\multicolumn{11}{c}{\textbf{Coverage} (The smaller, the better)}\\
\midrule
SMU & 7.67$\pm$0.21&7.81$\pm$0.20&7.81$\pm$0.32&7.84$\pm$0.18&7.72$\pm$0.18&7.56$\pm$0.24&7.76$\pm$0.20&7.62$\pm$0.26&7.75$\pm$0.07&\textbf{6.56$\pm$0.14}\\
G12EC & 4.80$\pm$0.09&4.97$\pm$0.09&4.83$\pm$0.08&4.85$\pm$0.10&4.79$\pm$0.11&4.87$\pm$0.10&4.86$\pm$0.16&4.88$\pm$0.09&4.70$\pm$0.08&\textbf{4.26$\pm$0.08}\\
PTB-XL & 3.27$\pm$0.05&3.30$\pm$0.07&3.27$\pm$0.04&3.36$\pm$0.03&3.28$\pm$0.06&3.28$\pm$0.06&3.29$\pm$0.06&3.29$\pm$0.05&3.27$\pm$0.05&\textbf{2.90$\pm$0.03}\\
Ningbo & 3.56$\pm$0.17&3.60$\pm$0.06&3.54$\pm$0.17&3.52$\pm$0.12&3.51$\pm$0.12&3.58$\pm$0.13&3.49$\pm$0.14&3.48$\pm$0.06&3.48$\pm$0.12&\textbf{2.97$\pm$0.08}\\
Chapman & 2.86$\pm$0.06&2.92$\pm$0.04&2.90$\pm$0.09&2.91$\pm$0.11&2.94$\pm$0.12&2.96$\pm$0.13&2.81$\pm$0.08&2.84$\pm$0.11&2.90$\pm$0.09&\textbf{2.61$\pm$0.06}\\
\midrule
\multicolumn{11}{c}{\textbf{MAP} (The greater, the better)}\\
\midrule
SMU & 52.89$\pm$1.70&52.44$\pm$1.17&52.70$\pm$2.25&52.91$\pm$1.55&53.18$\pm$1.20&53.55$\pm$1.76&53.11$\pm$1.87&53.34$\pm$1.45&53.52$\pm$1.35&\textbf{62.45$\pm$1.00}\\
G12EC & 39.82$\pm$1.22&39.67$\pm$0.97&39.59$\pm$0.96&38.60$\pm$0.70&39.92$\pm$1.05&39.74$\pm$1.40&39.91$\pm$0.98&40.34$\pm$0.99&40.73$\pm$0.63&\textbf{46.26$\pm$1.06}\\
PTB-XL & 41.53$\pm$0.99&41.85$\pm$1.03&41.40$\pm$0.91&38.93$\pm$0.85&42.39$\pm$0.62&40.98$\pm$1.19&41.84$\pm$0.23&41.31$\pm$1.24&42.18$\pm$0.64&\textbf{48.21$\pm$0.53}\\
Ningbo & 39.77$\pm$0.72&39.60$\pm$1.11&39.63$\pm$1.18&39.18$\pm$1.23&39.63$\pm$1.34&38.95$\pm$1.31&39.82$\pm$0.73&39.76$\pm$0.84&39.45$\pm$1.24&\textbf{47.13$\pm$0.67}\\
Chapman & 46.80$\pm$1.16&47.58$\pm$1.12&46.12$\pm$1.75&46.47$\pm$0.37&46.51$\pm$1.35&43.40$\pm$1.72&47.78$\pm$1.36&46.92$\pm$1.72&46.78$\pm$1.32&\textbf{51.05$\pm$1.10}\\
\midrule
\multicolumn{11}{c}{\textbf{Macro AUC} (The greater, the better)}\\
\midrule
SMU & 81.59$\pm$1.22&81.25$\pm$1.32&81.19$\pm$1.18&81.34$\pm$1.08&81.43$\pm$0.77&81.79$\pm$1.35&81.45$\pm$1.34&81.81$\pm$1.44&82.03$\pm$0.78&\textbf{86.36$\pm$0.46}\\
G12EC & 78.27$\pm$0.80&78.33$\pm$0.35&78.15$\pm$0.52&77.89$\pm$0.60&78.16$\pm$1.00&78.24$\pm$0.62&78.05$\pm$1.00&78.01$\pm$0.64&79.28$\pm$0.50&\textbf{81.65$\pm$0.16}\\
PTB-XL & 83.04$\pm$0.62&83.35$\pm$0.43&83.29$\pm$0.64&81.64$\pm$0.79&83.24$\pm$0.70&83.15$\pm$0.46&83.12$\pm$0.39&83.51$\pm$0.23&83.19$\pm$0.48&\textbf{87.10$\pm$0.42}\\
Ningbo & 86.39$\pm$1.20&86.53$\pm$1.21&86.46$\pm$1.30&86.21$\pm$0.96&86.85$\pm$1.01&85.78$\pm$1.58&86.55$\pm$1.22&86.97$\pm$0.60&87.15$\pm$0.87&\textbf{90.15$\pm$0.56}\\
Chapman & 85.11$\pm$1.03&85.12$\pm$0.72&85.25$\pm$1.07&84.05$\pm$1.56&85.21$\pm$0.67&83.63$\pm$1.19&85.57$\pm$0.74&85.36$\pm$1.28&84.73$\pm$0.90&\textbf{87.75$\pm$0.82}\\
\midrule
\multicolumn{11}{c}{\textbf{Macro F1 score} (The greater, the better)}\\
\midrule
SMU & 47.99$\pm$0.82&47.77$\pm$1.06&48.36$\pm$1.68&48.43$\pm$0.39&47.94$\pm$0.42&49.43$\pm$1.43&48.65$\pm$2.07&48.01$\pm$1.23&49.27$\pm$0.83&\textbf{56.79$\pm$0.89}\\
G12EC & 38.89$\pm$0.83&38.88$\pm$0.40&38.97$\pm$0.87&37.65$\pm$1.26&38.42$\pm$0.76&38.35$\pm$0.39&38.95$\pm$0.88&38.89$\pm$0.92&39.78$\pm$1.35&\textbf{44.36$\pm$1.30}\\
PTB-XL & 41.13$\pm$0.92&40.96$\pm$1.09&40.69$\pm$1.00&38.78$\pm$0.41&41.82$\pm$0.74&40.30$\pm$1.34&41.00$\pm$0.71&41.37$\pm$1.82&41.79$\pm$0.66&\textbf{47.69$\pm$0.24}\\
Ningbo & 40.13$\pm$0.81&39.64$\pm$0.77&40.02$\pm$1.07&39.23$\pm$0.68&39.27$\pm$1.04&39.17$\pm$1.18&40.31$\pm$0.79&39.89$\pm$0.62&39.60$\pm$0.92&\textbf{46.41$\pm$0.29}\\
Chapman & 46.57$\pm$0.76&45.93$\pm$1.25&45.73$\pm$1.34&45.71$\pm$1.23&46.72$\pm$1.16&43.59$\pm$1.24&46.86$\pm$0.98&45.70$\pm$1.70&46.16$\pm$1.40&\textbf{51.17$\pm$1.09}\\
\midrule
\multicolumn{11}{c}{\textbf{Macro $F_{\beta=2}$ score} (The greater, the better)}\\
\midrule
SMU & 58.19$\pm$1.25&57.40$\pm$1.81&58.21$\pm$1.60&57.62$\pm$1.65&58.09$\pm$0.89&58.30$\pm$1.94&57.83$\pm$2.14&58.00$\pm$1.43&58.50$\pm$0.98&\textbf{65.53$\pm$1.29}\\
G12EC & 49.88$\pm$1.17&49.09$\pm$0.35&49.63$\pm$0.63&48.84$\pm$1.56&49.62$\pm$0.86&48.80$\pm$0.66&49.29$\pm$0.85&49.45$\pm$0.67&50.85$\pm$1.02&\textbf{54.54$\pm$0.99}\\
PTB-XL & 50.75$\pm$1.06&51.04$\pm$0.76&50.44$\pm$0.91&48.18$\pm$0.51&50.94$\pm$0.80&50.11$\pm$1.57&50.12$\pm$0.04&50.89$\pm$1.17&50.48$\pm$0.83&\textbf{56.30$\pm$0.22}\\
Ningbo & 48.95$\pm$0.61&48.21$\pm$0.97&48.71$\pm$1.31&47.91$\pm$1.01&48.37$\pm$1.21&48.07$\pm$1.85&48.86$\pm$1.06&48.68$\pm$0.75&48.77$\pm$1.05&\textbf{55.39$\pm$0.51}\\
Chapman & 54.09$\pm$0.69&53.57$\pm$1.42&53.25$\pm$2.01&53.45$\pm$1.29&54.21$\pm$1.52&50.63$\pm$0.68&54.35$\pm$0.70&52.82$\pm$1.59&53.28$\pm$1.17&\textbf{58.58$\pm$0.67}\\
\bottomrule
\end{tabular*}
}
\end{center}
\end{extendtable*}

\begin{extendtable*}[h]
\fontsize{7}{11}\selectfont
\setlength{\tabcolsep}{0.1em}
\begin{center}
\caption{Cross-device external validation of LiteHeart and the SOTA methods with 10\% labeled samples. The mean performance and standard deviations on 5 databases are shown across 4 seeds.}
\label{tab:external_tiny10}
\scalebox{1}{
\color{black}
\begin{tabular*}{\hsize}{@{}@{\extracolsep{\fill}}lccccccccccc@{}}
\toprule
Methods & KD &DKD & CC & RKD & DIST & PKT & PSM & WKD  & MVKT & \textbf{LiteHeart} \\
\midrule
Ranking loss & 0.16$\pm$0.00&0.16$\pm$0.01&0.16$\pm$0.01&0.17$\pm$0.01&0.16$\pm$0.01&0.17$\pm$0.01&0.16$\pm$0.01&0.17$\pm$0.00&0.16$\pm$0.01&\textbf{0.13$\pm$0.00}\\
Coverage & 5.88$\pm$0.12&5.85$\pm$0.09&5.78$\pm$0.11&6.07$\pm$0.12&5.98$\pm$0.17&6.00$\pm$0.14&5.82$\pm$0.17&5.94$\pm$0.10&5.81$\pm$0.10&\textbf{5.21$\pm$0.03}\\
MAP & 47.39$\pm$1.14&48.13$\pm$1.14&48.70$\pm$1.27&45.50$\pm$1.57&48.09$\pm$0.54&46.18$\pm$3.40&48.16$\pm$1.03&47.98$\pm$0.81&48.64$\pm$1.50&\textbf{54.53$\pm$0.83}\\
Macro AUC & 81.69$\pm$0.75&81.47$\pm$0.66&81.06$\pm$0.91&78.61$\pm$1.71&81.61$\pm$0.54&79.89$\pm$2.58&81.13$\pm$0.70&81.55$\pm$0.87&81.85$\pm$1.24&\textbf{85.44$\pm$0.70}\\
Macro F1& 43.35$\pm$1.80&43.94$\pm$1.35&43.16$\pm$2.68&40.92$\pm$2.62&44.12$\pm$2.14&41.17$\pm$4.36&45.07$\pm$1.83&43.56$\pm$0.44&43.96$\pm$1.30&\textbf{51.74$\pm$0.59}\\
Macro $F_{\beta=2}$& 50.63$\pm$1.35&50.97$\pm$1.25&51.99$\pm$2.23&49.40$\pm$2.39&50.26$\pm$1.83&48.53$\pm$3.84&52.42$\pm$1.71&51.11$\pm$1.64&50.47$\pm$2.19&\textbf{58.30$\pm$0.91}\\
\bottomrule
\end{tabular*}
}
\end{center}
\end{extendtable*}

\begin{extendtable*}[h]
\fontsize{7}{11}\selectfont
\setlength{\tabcolsep}{0.1em}
\begin{center}
\caption{Cross-device external validation of LiteHeart and the SOTA methods with 5\% labeled samples. The mean performance and standard deviations on 5 databases are shown across 4 seeds.}
\label{tab:external_tiny05}
\scalebox{1}{
\color{black}
\begin{tabular*}{\hsize}{@{}@{\extracolsep{\fill}}lccccccccccc@{}}
\toprule
Methods & KD &DKD & CC & RKD & DIST & PKT & PSM & WKD  & MVKT & \textbf{LiteHeart} \\
\midrule
Ranking loss & 0.17$\pm$0.01&0.17$\pm$0.01&0.18$\pm$0.01&0.19$\pm$0.01&0.18$\pm$0.01&0.18$\pm$0.01&0.17$\pm$0.01&0.17$\pm$0.01&0.17$\pm$0.00&\textbf{0.14$\pm$0.01}\\
Coverage & 6.08$\pm$0.15&6.00$\pm$0.20&6.16$\pm$0.21&6.27$\pm$0.23&6.19$\pm$0.17&6.19$\pm$0.20&6.16$\pm$0.23&6.02$\pm$0.21&6.11$\pm$0.06&\textbf{5.41$\pm$0.18}\\
MAP & 43.97$\pm$1.71&44.79$\pm$1.30&43.17$\pm$1.36&40.85$\pm$1.62&43.12$\pm$0.49&42.84$\pm$1.13&42.96$\pm$1.26&45.67$\pm$0.76&45.12$\pm$0.61&\textbf{51.69$\pm$0.94}\\
Macro AUC & 79.32$\pm$1.28&80.02$\pm$1.02&78.86$\pm$1.25&76.62$\pm$2.00&78.86$\pm$0.35&78.63$\pm$1.37&78.76$\pm$0.82&80.78$\pm$0.77&80.17$\pm$0.36&\textbf{83.92$\pm$1.01}\\
Macro F1& 40.65$\pm$2.15&40.95$\pm$1.77&39.95$\pm$0.58&38.58$\pm$2.11&39.29$\pm$0.88&38.55$\pm$0.69&39.82$\pm$2.32&41.71$\pm$2.12&40.61$\pm$1.51&\textbf{47.36$\pm$1.36}\\
Macro $F_{\beta=2}$& 46.31$\pm$1.80&46.59$\pm$1.77&46.11$\pm$1.77&44.32$\pm$2.18&45.70$\pm$1.23&45.36$\pm$0.18&46.30$\pm$1.70&49.15$\pm$1.35&46.85$\pm$2.48&\textbf{53.65$\pm$1.55}\\
\bottomrule
\end{tabular*}
}
\end{center}
\end{extendtable*}

\begin{extendtable*}[h]
\fontsize{7}{11}\selectfont
\setlength{\tabcolsep}{0.1em}
\begin{center}
\caption{Cross-device external validation of LiteHeart and the SOTA methods with 2.5\% labeled samples. The mean performance and standard deviations on 5 databases are shown across 4 seeds.}
\label{tab:external_tiny025}
\scalebox{1}{
\color{black}
\begin{tabular*}{\hsize}{@{}@{\extracolsep{\fill}}lccccccccccc@{}}
\toprule
Methods & KD &DKD & CC & RKD & DIST & PKT & PSM & WKD  & MVKT & \textbf{LiteHeart} \\
\midrule
Ranking loss & 0.18$\pm$0.01&0.18$\pm$0.01&0.18$\pm$0.01&0.18$\pm$0.01&0.18$\pm$0.01&0.18$\pm$0.01&0.18$\pm$0.01&0.17$\pm$0.01&0.17$\pm$0.01&\textbf{0.15$\pm$0.01}\\
Coverage & 6.14$\pm$0.10&6.27$\pm$0.19&6.18$\pm$0.17&6.20$\pm$0.27&6.10$\pm$0.30&6.19$\pm$0.26&6.19$\pm$0.06&6.06$\pm$0.09&6.04$\pm$0.14&\textbf{5.61$\pm$0.16}\\
MAP & 41.80$\pm$1.75&40.85$\pm$1.24&41.41$\pm$0.81&39.90$\pm$2.33&42.31$\pm$1.05&41.02$\pm$2.22&41.39$\pm$0.74&42.43$\pm$1.35&43.02$\pm$2.54&\textbf{49.38$\pm$0.79}\\
Macro AUC & 79.41$\pm$0.94&79.37$\pm$0.84&79.20$\pm$0.42&77.14$\pm$1.73&79.39$\pm$0.53&78.31$\pm$1.68&79.12$\pm$0.49&80.09$\pm$0.77&80.40$\pm$1.18&\textbf{83.67$\pm$0.85}\\
Macro F1& 37.32$\pm$1.21&37.49$\pm$1.87&37.96$\pm$0.61&34.99$\pm$2.83&37.56$\pm$2.54&37.76$\pm$2.75&38.65$\pm$1.66&39.71$\pm$1.40&37.29$\pm$0.91&\textbf{45.14$\pm$0.98}\\
Macro $F_{\beta=2}$& 44.03$\pm$2.15&44.36$\pm$2.08&43.98$\pm$1.44&42.04$\pm$1.74&44.38$\pm$2.87&45.33$\pm$3.17&45.24$\pm$0.88&46.66$\pm$1.07&45.35$\pm$2.11&\textbf{51.99$\pm$1.39}\\
\bottomrule
\end{tabular*}
}
\end{center}
\end{extendtable*}

\end{document}